\documentclass[preprint]{aastex62}

\graphicspath{{./}{figures/}}

\received{September 30, 2020}
\revised{November 30, 2020}
\accepted{December 1, 2020}
\submitjournal{PSJ - A'Hearn Symposium Focus Issue}

\shorttitle{Comet Wirtanen Rotation}
\shortauthors{Farnham et al.}

\begin{document}

\title{Narrowband Observations of Comet 46P/Wirtanen During Its Exceptional
  Apparition of 2018/19 I: Apparent Rotation Period and Outbursts}

\correspondingauthor{Tony L. Farnham}
\email{farnham@astro.umd.edu}

\author[0000-0002-4767-9861]{Tony L. Farnham}
\affiliation{University of Maryland,
Department of Astronomy,
College Park, MD 20742, USA}

\author[0000-0003-2781-6897]{Matthew M. Knight}
\affiliation{Department of Physics, 
United States Naval Academy, 572C Holloway Rd,
Annapolis, MD 21402, USA}
\affiliation{University of Maryland,
Department of Astronomy,
College Park, MD 20742, USA}

\author{David G. Schleicher}
\affiliation{Lowell Observatory,
1400 W. Mars Hill Rd.,
Flagstaff, AZ 86001, USA}

\author[0000-0002-4230-6759]{Lori M. Feaga}
\affiliation{University of Maryland,
Department of Astronomy,
College Park, MD 20742, USA}

\author[0000-0002-2668-7248]{Dennis Bodewits}
\affiliation{Department of Physics, 
Leach Science Center, 
Auburn University, Auburn, AL, 36832, USA}

\author{Brian A. Skiff}
\affiliation{Lowell Observatory,
1400 W. Mars Hill Rd.,
Flagstaff, AZ 86001, USA}

\author{Josephine Schindler}
\affiliation{Lowell Observatory,
1400 W. Mars Hill Rd.,
Flagstaff, AZ 86001, USA}



\begin{abstract}

We obtained broadband and narrowband images of the hyperactive comet
46P/Wirtanen on 33~nights during its 2018/2019 apparition, when the comet
made an historic close approach to the Earth.  With our extensive coverage,
we investigated the temporal behavior of the comet on both seasonal and
rotational timescales.  CN observations were used to explore the coma
morphology, revealing that there are two primary active areas that produce
spiral structures.  The direction of rotation of these structures changes
from pre- to post-perihelion, indicating that the Earth crossed the comet's
equatorial plane sometime around perihelion.  We also
used the CN images to 
create photometric lightcurves that consistently show two peaks in the
activity, confirming the two source regions.  We measured the nucleus'
apparent rotation period at a number of epochs using both the morphology and
the lightcurves. These results all show that the rotation period is
continuously changing throughout our observation window, increasing from
8.98~hr in early November to 9.14~hr around perihelion and then
decreasing again to 8.94~hr in February. Although the geometry changes
rapidly around perihelion, the period changes cannot primarily be due to
synodic effects.  The repetition of structures in the coma, both within a
night and from night-to-night, strongly suggests the nucleus is in a
near-simple rotation state.  We also detected two outbursts,
one on December~12 and the other on January~28.  Using apparent velocities of
the ejecta in these events, 68$\pm$5~m~s$^{-1}$ and 162$\pm$15~m~s$^{-1}$,
respectively, we derived start times of 2018~December~12 at
00:13~UT~$\pm$7~min and 2019~January~27 at 20:01~UT~$\pm$30~min.

\end{abstract}



\section{Introduction} \label{sec:intro}

Comet 46P/Wirtanen is a Jupiter family comet that was discovered on
1948~January~17 by Carl Wirtanen at the Lick Observatory.  Its orbit is such
that it frequently gets close enough to Jupiter to be perturbed, and this
has happened several times in the last century.  In 1972, Wirtanen's
perihelion distance decreased from 1.61~au to 1.26~au, and then again in 1984 it
dropped to 1.06~au, where it currently remains.  It is not known if
this is the comet's closest foray to the Sun, but in the past few orbits, it
has been experiencing more intense heating than it has for some time.

Wirtanen's current orbit is readily accessible, making it desirable for
spacecraft missions.  Despite the fact that very little was known about the
comet, it was selected as the target for the Rosetta mission in 1994
\citep{ESA:Rosetta}.  Although observations were obtained in support of this
mission, conditions were poor during the 1997 and 2002 apparitions, so
additional understanding was somewhat limited.  In 2003, due to delays in the
Rosetta launch date, Wirtanen was dropped as the target.  Wirtanen was
later selected as the target of the proposed Comet Hopper Discovery mission
(2011 Phase A study, unselected), and has been the target of several other
proposed missions.  The fact that it is repeatedly considered as a target
suggests that there is a strong chance that it will be visited in the future,
and understanding its physical characteristics and behavior would help to
reduce the costs and risks involved in the design and planning of any
mission.

In addition to being a candidate for a spacecraft target, Wirtanen is an
interesting object in its own right.  It has a relatively small nucleus, with
an effective radius of 0.60~km and axial ratio $\ge1.4$
\citep{LamyEtal:wirthst,Boehn:wirtanen}. Given its water production rate,
$\sim1.5\times10^{28}$~molec/s \citep{FarnhamSchl:wirtphot,
  GroussinEtal:wirtact, KobayashiEtal:wirtanen, CombiEtal:water_survey},
Wirtanen is a hyperactive comet, emitting more water than would be expected,
based on its size and standard water vaporization models
\citep{CowanAhearn:vapor}.  The Deep Impact eXtended Investigation, which
visited another hyperactive comet, 103P/Hartley~2, showed that this
hyperactivity was produced by icy grains that were dragged into the coma by
CO$_2$ emission \citep{AhearnEtal:epoxi,ProtoEtal:H2_icegrains}.  In many
respects, Wirtanen and Hartley~2 are comparable and comparisons between them
could provide insight into the family of hyperactive comets.

Wirtanen's 2018/2019 apparition provided the first excellent opportunity to
investigate the comet in detail.  Only four days after its December~12
perihelion, the comet made an historically close approach, passing only
0.0775~au (30 lunar distances) from the Earth.  With spatial scales as small
as 57~km/arcsec and the quality of ground-based telescopes, this
offered conditions similar to those that would be seen in a distant flyby,
while allowing numerous ground-based telescopes, using instruments that could
never be carried on a spacecraft, to study the inner coma of the comet.
Because the comet was near opposition during its apparition, it was
observable for many hours during the night, allowing long-term monitoring for
months during the event.

We took this opportunity to obtain narrowband filter images of the comet on
33 nights (in 9 observing runs, plus occasional sampling with a robotic
telescope) spanning close approach, to characterize the comet's behavior.  In
this paper, we present analyses of these data, using using both morphology
and photometric measurements to explore the comet's seasonal and rotational
characteristics.  We assume that these changes are the result of variability
in the CN production as the nucleus rotates (short term) and changes its
orientation with respect to the Sun and Earth (long term).  Under this
assumption, we use both the photometric lightcurve and the repeatability of
features in the coma to derive the instantaneous rotation period of the
nucleus and to look for changes throughout the comet's perihelion passage.
In a companion paper \citep{KnightEtal:wirtanen}, we used Monte-Carlo models
of the coma structure to derive the orientation of the spin axis and the
locations of any active areas, as well as resolving the extent to which
synodic effects can affect the perceived rotation period.

\section{Observations and Data Reduction}

\subsection{Observations} \label{sec:obs}

The majority of the data used in this work were obtained at the 4.3-m Lowell
Discovery Telescope (LDT; formerly Discovery Channel Telescope) and the
Lowell Observatory John S. Hall 42-inch (1.1-m) Telescope.  We also obtained
images at the robotic Lowell 31-inch (0.8-m) telescope, but these tend to be
isolated ``snapshot'' observations.  By themselves, the 31-inch data are not
suitable for period determination, but we did make use of several nights to
extend the temporal baseline of some of the more complete sequences.
Specific nights and the relevant geometric conditions for images used in this
paper are listed in Table~\ref{tab:obs}.  Images from the LDT were obtained
with the Large Monolithic Imager (LMI), which has a 6.1k$\times$6.1k e2v CCD
with a 12.3-arcmin field of view.  On-chip 2$\times$2 binning produces a
pixel scale of 0.24~arcsec.  Images from the Hall telescope were obtained
with a 4k$\times$4k e2v CCD231-84 chip, with 2$\times$2 binned pixels of
0.74~arcsec (though 2019 January 4 was binned 3$\times$3 for 1.1~arcsec
pixels), and 31-in images were obtained with a 2k$\times$2k e2v CCD42-40 chip
with unbinned 0.46-arcsec pixels.  On all telescopes, we
used a broadband R (or r$^\prime$) filter, as well as HB narrowband comet
filters \citep{FarnhamEtal:hbfilters}.  The narrowband filters isolate five
different gas species (OH, NH, CN, C$_3$, and C$_2$) and several different
continuum bands.  We obtained different combinations of filters on different
nights, depending on observing conditions, etc., though broadband R and CN
filters were used to monitor the comet's morphology and obtained as
frequently as possible.  Exposure times were 120--300~s for~CN and 5--120~s
for~R.  Whenever possible, sets of three to five images were obtained in
sequence, allowing us to later combine them using a median filter to improve
the S/N and reduce the interference from cosmic rays and background stars.
In this work, we primarily focus on the CN observations, and will address the
other gas species and continuum images in the companion paper by
\cite{KnightEtal:wirtanen}.

\begin{deluxetable*}{llrrclcccrl}
\renewcommand{\baselinestretch}{0.8}
\tablecaption{Observing Circumstances\tablenotemark{a} \label{tab:obs}}
\tablecolumns{11}
\tablewidth{0pt}
\tablehead{
\colhead{Date} &
\colhead{UT Range} &
\colhead{Dur.\tablenotemark{b}} &
\colhead{$\Delta t$\tablenotemark{c}} &
\colhead{Phase} & 
\colhead{Tel.\tablenotemark{e}} & 
\colhead{$r_H$\tablenotemark{f}} &
\colhead{$\Delta$\tablenotemark{g}} & 
\colhead{$\alpha$\tablenotemark{h}} & 
\colhead{PA$_\sun$\tablenotemark{i}} & 
\colhead{Quality} \\
\colhead{} & \colhead{} & \colhead{(hr)} & \colhead{(day)} & 
\colhead{Groups\tablenotemark{d}} & \colhead{} & \colhead{(au)} &
\colhead{(au)} & \colhead{(deg)} &  \colhead{(deg)} & \colhead{} }
\startdata
2018 Nov 01 &  04:37--08:38 &  4.02 &--41.65 & 1,A   & 42in & 1.193 & 0.273 & 38.0 & 188.1 & P. Cloudy \\
2018 Nov 02 &  04:30--08:55 &  4.42 &--40.65 & 1,A   & 42in & 1.187 & 0.267 & 38.6 & 189.2 & Cirrus \\
2018 Nov 03 &  04:21--08:53 &  4.53 &--39.65 & 1,A   & LDT  & 1.181 & 0.262 & 39.1 & 190.3 & P. Cloudy \\ 
2018 Nov 04 &  04:13--08:34 &  4.35 &--38.66 & 1,A   & 42in & 1.176 & 0.256 & 39.7 & 191.4 & Cirrus \\
2018 Nov 09 &  04:36--07:48 &  3.20 &--33.67 & 2,A   & 42in & 1.148 & 0.229 & 42.3 & 196.6 & Clear  \\
2018 Nov 11 &  04:16--08:01 &  3.75 &--31.67 & 2,A   & 42in & 1.138 & 0.219 & 43.2 & 198.6 & Clear  \\
2018 Nov 12 &  04:15--07:26 &  3.18 &--30.69 & 2,A   & 42in & 1.133 & 0.214 & 43.7 & 199.5 & Clear  \\
2018 Nov 13 &  04:34--07:45 &  3.18 &--29.67 & 2,A   & 42in & 1.128 & 0.208 & 44.1 & 200.5 & Clear  \\
2018 Nov 26 &  04:49, 06:11 &  ---  &--16.70 &  B    & 31in & 1.079 & 0.143 & 46.7 & 211.3 & P. Cloudy \\
2018 Nov 27 &  04:43, 06:08 &  ---  &--15.71 &  B    & 31in & 1.077 & 0.138 & 46.5 & 212.1 & Clear  \\
2018 Nov 29 &  04:40, 06:05 &  ---  &--13.69 &  B    & 31in & 1.072 & 0.128 & 45.8 & 213.7 & P. Cloudy \\
2018 Dec 03 &  02:06--09:05 &  6.98 & --9.70 & 3,B,C & LDT  & 1.064 & 0.111 & 43.3 & 217.3 & Cirrus \\
2018 Dec 04 &  01:55--08:41 &  6.77 & --8.71 & 3,B,C & LDT  & 1.062 & 0.107 & 42.3 & 218.3 & Cirrus \\
2018 Dec 06 &  01:36--06:57 &  5.35 & --6.75 & 3,B,C & LDT  & 1.059 & 0.099 & 39.9 & 220.7 & P. Cloudy \\  
2018 Dec 09 &  01:46--02:07 &  0.35 & --3.85 & C     & 42in & 1.057 & 0.089 & 35.2 & 225.7 & Clear  \\
2018 Dec 10 &  00:46--10:11 &  9.41 & --2.70 & C     & 42in & 1.056 & 0.086 & 33.0 & 228.5 & Cirrus \\
2018 Dec 12 &  01:47--07:10 &  5.38 & --0.74 & ---   & LDT  & 1.055 & 0.082 & 28.9 & 234.9 & Cirrus \\
2018 Dec 13 &  01:15--07:25 &  6.17 &   0.25 & 4     & LDT  & 1.055 & 0.080 & 26.7 & 239.2 & Clear  \\
2018 Dec 14 &  01:08--07:15 &  6.12 &   1.25 & 4,5   & LDT  & 1.055 & 0.079 & 24.5 & 244.6 & Clear  \\
2018 Dec 15 &  01:32--09:49 &  8.28 &   2.31 & 4,5,6 & LDT  & 1.056 & 0.078 & 22.3 & 251.7 & P. Cloudy \\
2018 Dec 16 &  01:40--11:26 &  9.77 &   3.35 & 5,6   & LDT  & 1.056 & 0.077 & 20.5 & 260.2 & Clear  \\
2018 Dec 17 &  01:30--11:18 &  9.80 &   4.34 & 6     & LDT  & 1.057 & 0.078 & 19.2 & 239.7 & P. Cloudy \\
2018 Dec 27 &  03:36        &  ---  &  14.42 & D     & 31in & 1.073 & 0.102 & 26.6 & 352.5 & P. Cloudy \\
2018 Dec 30 &  02:41, 13:19 &  ---  &  17.40 & D     & 31in & 1.081 & 0.115 & 29.6 &   3.5 & Cirrus \\
2018 Dec 31 &  04:23        &  ---  &  18.26 & D     & 31in & 1.084 & 0.119 & 30.2 &   6.1 & P. Cloudy \\
2019 Jan 03 &  01:45--13:12 & 11.45 &  21.38 & 7,D,E & 42in & 1.094 & 0.134 & 32.1 &  13.7 & Clear  \\
2019 Jan 04 &  04:14--13:12 &  8.97 &  22.44 & 7,D,E & 42in & 1.098 & 0.139 & 32.5 &  15.6 & Clear  \\
2019 Jan 12 &  01:35--13:30 & 11.92 &  30.39 & E,F   & LDT  & 1.131 & 0.182 & 33.4 &  23.3 & P. Cloudy \\
2019 Jan 26 &  02:01--13:10 & 11.15 &  44.39 & 8,F,G & 42in & 1.210 & 0.273 & 30.5 &  17.5 & P. Cloudy \\
2019 Jan 27 &  02:01--06:40 &  4.65 &  45.25 & 8,F,G & 42in & 1.215 & 0.279 & 30.3 &  16.6 & P. Cloudy \\
2019 Jan 28 &  02:00--13:07 & 11.12 &  46.39 & 8,F,G & 42in & 1.223 & 0.287 & 30.0 &  15.4 & Cirrus \\
2019 Feb 08 &  04:00--13:30 &  9.50 &  57.44 & 9,G   & 42in & 1.300 & 0.371 & 27.5 &   1.4 & Cirrus \\
2019 Feb 09 &  02:33--05:33 &  3.00 &  58.24 & 9,G   & 42in & 1.305 & 0.376 & 27.4 &   0.5 & P. Cloudy \\
\enddata
\tablenotetext{a}{ Parameters listed are for the midtime of each night's observations}
\tablenotetext{b}{ Nightly duration of the observations}
\tablenotetext{c}{ Time from perihelion}
\tablenotetext{d}{ Groups used to phase data in the morphology analyses;
  Numbers link nights combined over a single observing run, 
  Letters combine nights over two runs}  
\tablenotetext{e}{ Telescope: LDT--Lowell Discovery Telescope;
  42in--42-in Hall Telescope; 31in--Lowell 31-in Telescope }
\tablenotetext{f}{ Heliocentric distance}
\tablenotetext{g}{ Geocentric distance}
\tablenotetext{h}{ Solar phase angle}
\tablenotetext{i}{ Position angle of the Sun}
\end{deluxetable*}
\renewcommand{\baselinestretch}{1.0}

\subsection{Data Reduction} \label{sec:reduc}

We used standard reduction procedures for bias removal and flat fielding.
Usually, the continuum underlying the CN images is minimal, so for our
analyses of the coma morphology, photometric calibration of the images was
unnecessary and was not done for this work (see Section~\ref{sec:features}).
This allowed us to use images from all nights listed, including those
obtained under non-photometric conditions.  Individual images were then
registered on their optocenter using a 2D Gaussian fit to the innermost coma
to determine their centroid.  After registration, the sets of three to five
images that were obtained together (Section~\ref{sec:obs}) were combined into
the ``final'' images that were used in our subsequent analyses.

After data reduction and enhancement (Section~\ref{sec:enhance}), we rescaled
the images and trimmed them to a common physical scale, to enable direct
comparisons.  Images obtained between 2018~November~17 and 2019~January~6
($\pm25$~days from perihelion) were trimmed to a 30,000~km field-of-view,
while data obtained outside of this window were trimmed to 60,000~km.

\subsection{Image Enhancement} \label{sec:enhance}

As with many comets, the CN coma of comet Wirtanen is fairly symmetric when
viewed in unprocessed images, but exhibits a wealth of morphological detail
when image enhancements are applied.  In this work, we adopted three
different techniques \citep{SchlFarn:comets2,SamarLarson:imenhance}, each of
which removes the bulk falloff in a different manner.  These techniques
reveal different aspects of the coma that are used to explore the comet's
temporal behavior.  For our first enhancement technique, we computed the
azimuthally averaged radial profile in each image and divided it out to
remove the bulk shape of the coma.  This is a relatively benign technique
that minimizes artifacts and centroiding uncertainties, while preserving the
relative brightness asymmetries in different directions.

Our second technique takes advantage of the fact that we have excellent
temporal coverage of the comet in most of our observing runs.  This
enhancement uses a temporally-averaged mask that is applied to all nights
from a given observing run.  For each run, we selected a sequence of 8--10
frames at roughly equal intervals of rotational phase (using a 9.0-hr period
here).  These were then scaled and averaged together to produce a
temporally-averaged master frame that smoooths out the coma variations over a
full rotation period.  This mask was divided out of each individual image in
the group.  This enhancement is particularly powerful for several reasons: it
applies the same mask to each individual frame, providing a uniform
enhancement; it removes the majority of the coma, revealing faint features
that are lost in the brightness gradients that are retained in other
techniques; and the features that remain are those that change with rotation
making it a valuable tool for determining image phasing over several nights.
It is especially valuable for revealing features on the darker side of the
coma, which are often lost in contrast to the bright side. On the other hand,
this technique also has drawbacks.  Because it aggressively removes the bulk
of the coma, low-level morphologies are revealed, and the detailed appearance
of the features can be sensitive to seeing variations and uncertanties in the
background removal, and the region near the optocenter can also be sensitive
to uncertainties in the centroiding. Most importantly, because only the
rotational changes in the coma are retained, the apparent morphology does not
necessarily represent the true shapes of the outflowing jet material, which
can be misleading if the results are not interpreted in conjunction with
other enhancements.

Our third enhancement technique is a combination of the first two, in which
we derived the average radial profile of the temporally averaged mask, and
then divided that profile out of each individual frame.  The result is similar
to that from our first enhancement, but it provides a check that purely
azimuthal features are not being lost, as could be the case when the
azimuthal average profile is derived from each individual image.

\begin{figure}[hbt!]
\epsscale{0.75}
\plotone{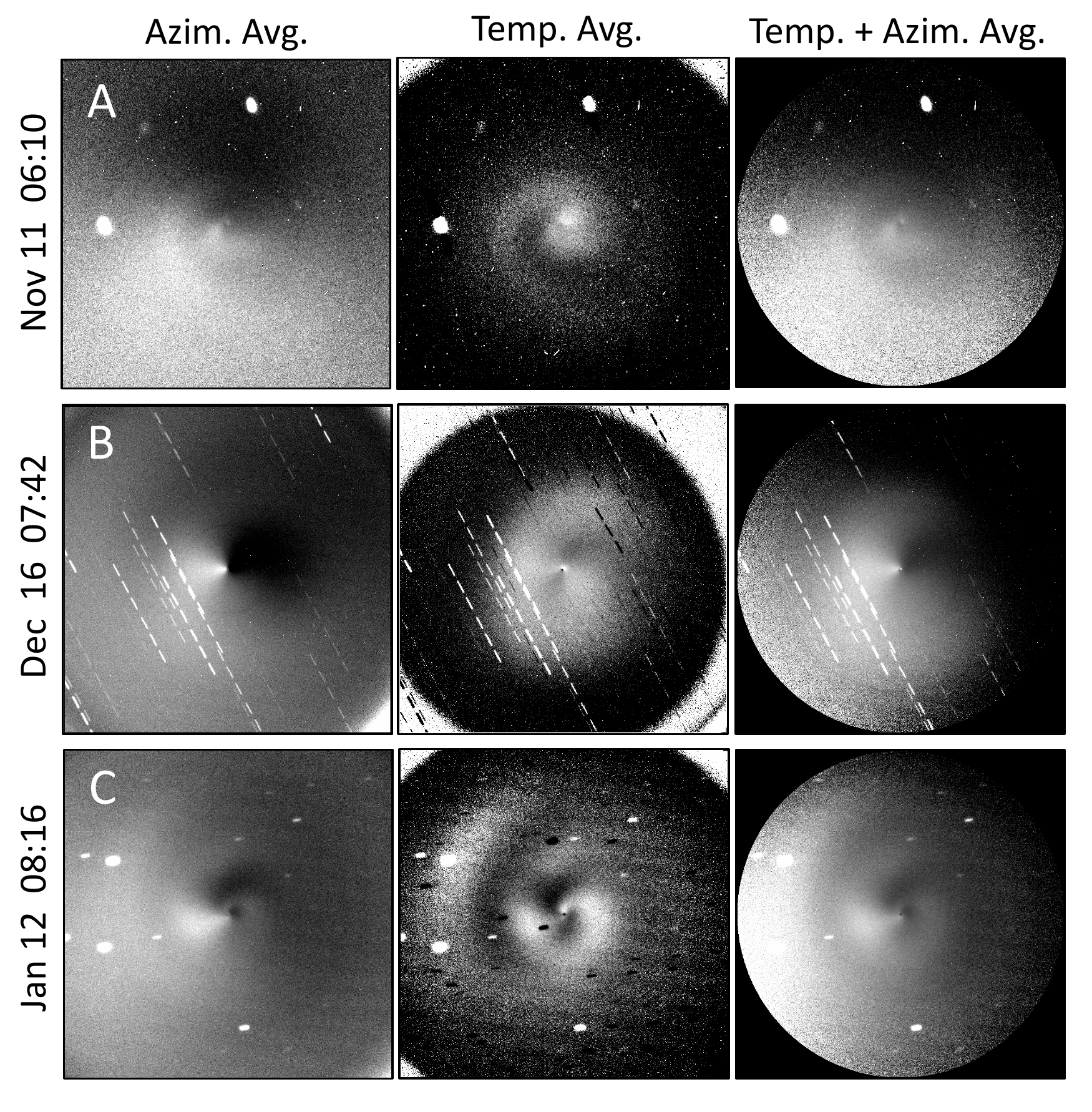}
\caption{Examples of image enhancement techniques applied to sample images
  from three dates around perihelion.  This figure also illustrates some of
  the basic morphology seen around perihelion, with two jets approximately
  180\degr\ apart arising from the nucleus on each date.  Images in
  rows~A and~C are 60,000~km across, while those in row~B are 30,000~km
  across.  North is up and East is to the left. The bright streaks are star
  trails, while the dark streaks in the rotational averaged column are the
  negative values produced by stars from the other temporally-combined
  images. \label{fig:imenh}}
\end{figure}

Figure~\ref{fig:imenh} shows the results from the three enhancement
techniques as applied to CN images from three different dates, demonstrating
how each technique reveals different aspects of the coma morphology.  It
illustrates that the azimuthally averaged versions are better at retaining
the true shapes of features as well as the basic brightness asymmetries in
different directions.  In contrast, the temporally averaged enhancement
removes the asymmetries, which more clearly highlights fainter features, but
can also alter the apparent shapes of structures.  

When comparing the coma morphology in the enhanced images, we find that the
primary features remained consistent over multiple rotations (aside from
changes in the viewing geometry).  However, subtle, low-level features (e.g.,
faint arcs near the edge of the field) can sometimes exhibit notable
differences, due to the effects of seeing and transparency variations,
background sky removal, and even contrast display levels.  Thus, caution
should be taken in interpreting the differences in these low-level features.

\section{Coma Morphology and Rotational Analysis \label{sec:morphology}} 

\subsection{Feature Descriptions and Motions \label{sec:features}}

\begin{figure}[hbt!]
\epsscale{0.6}
\plotone{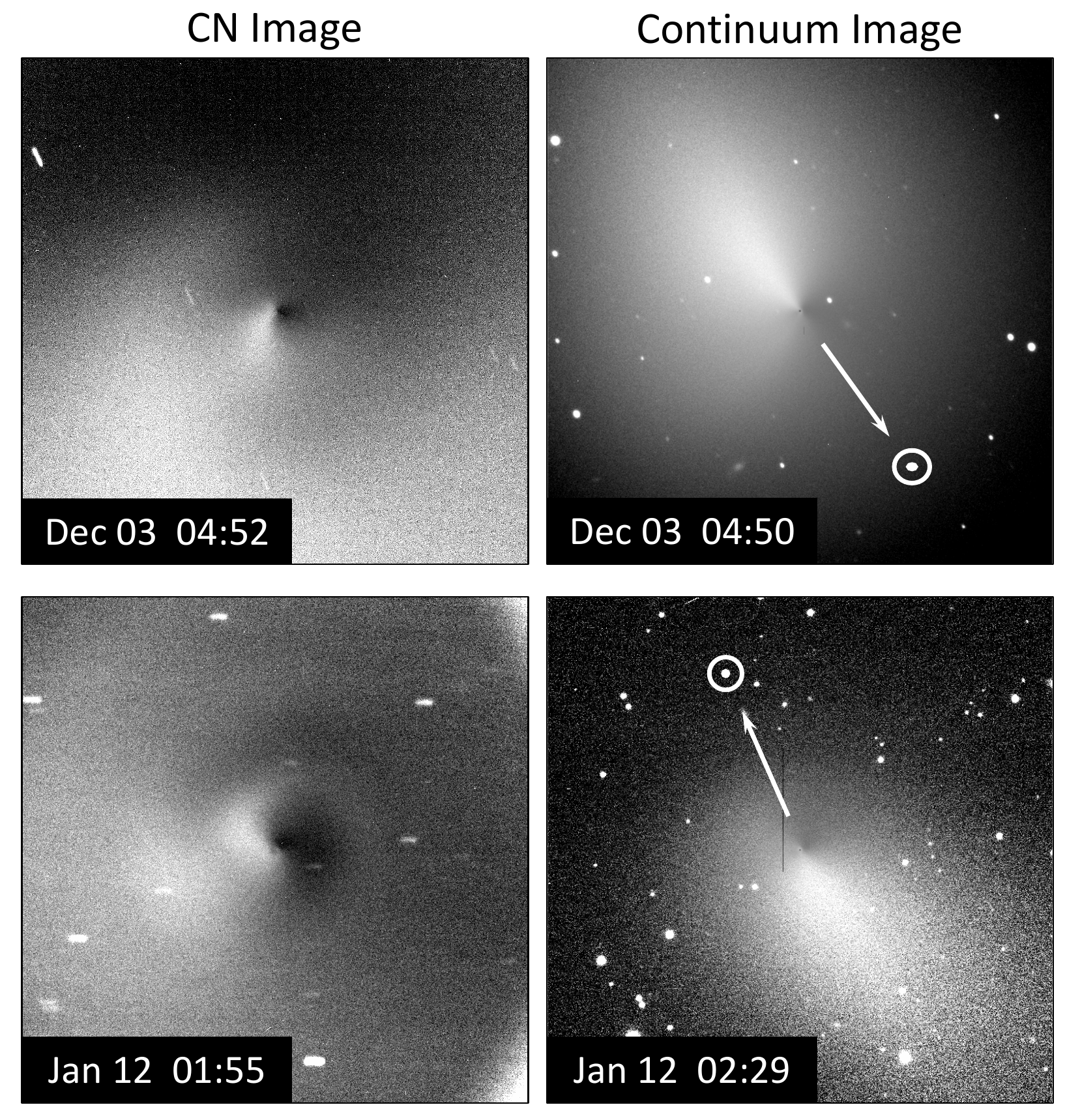}
\caption{Comparison of sample CN and continuum images, showing that the dust
  features are usually not detected in the CN morphology. CN images are
  enhanced by division of an azimuthal averaged profile and the dust images
  are enhanced by division of a 1/$\rho$ profile. December images have a field of
  view of 30,000~km, January images of 60,000~km.
  \label{fig:CN_Dust}}  
\end{figure}

We explored the dust morphology to determine how it might affect our CN
analyses.  On most observing runs, the typical underlying continuum is not
significant (typically less than a few percent) when compared to the CN.
Near close approach, the concentrations of dust near the optocenter can be
detected in some of the enhanced CN images (see \cite{KnightEtal:wirtanen}
for more details).  Fortunately, the dust morphology differs from the CN
morphology (Figure~\ref{fig:CN_Dust}) and essentially remains unchanging with
rotation.  Thus, when dust is detected, it should not affect our search for
periodicity.  An outburst detected on December 12, discussed further in
Section~\ref{sec:outbursts}, is one exception.

\begin{figure}[hbt!]
\epsscale{1.0}
\plotone{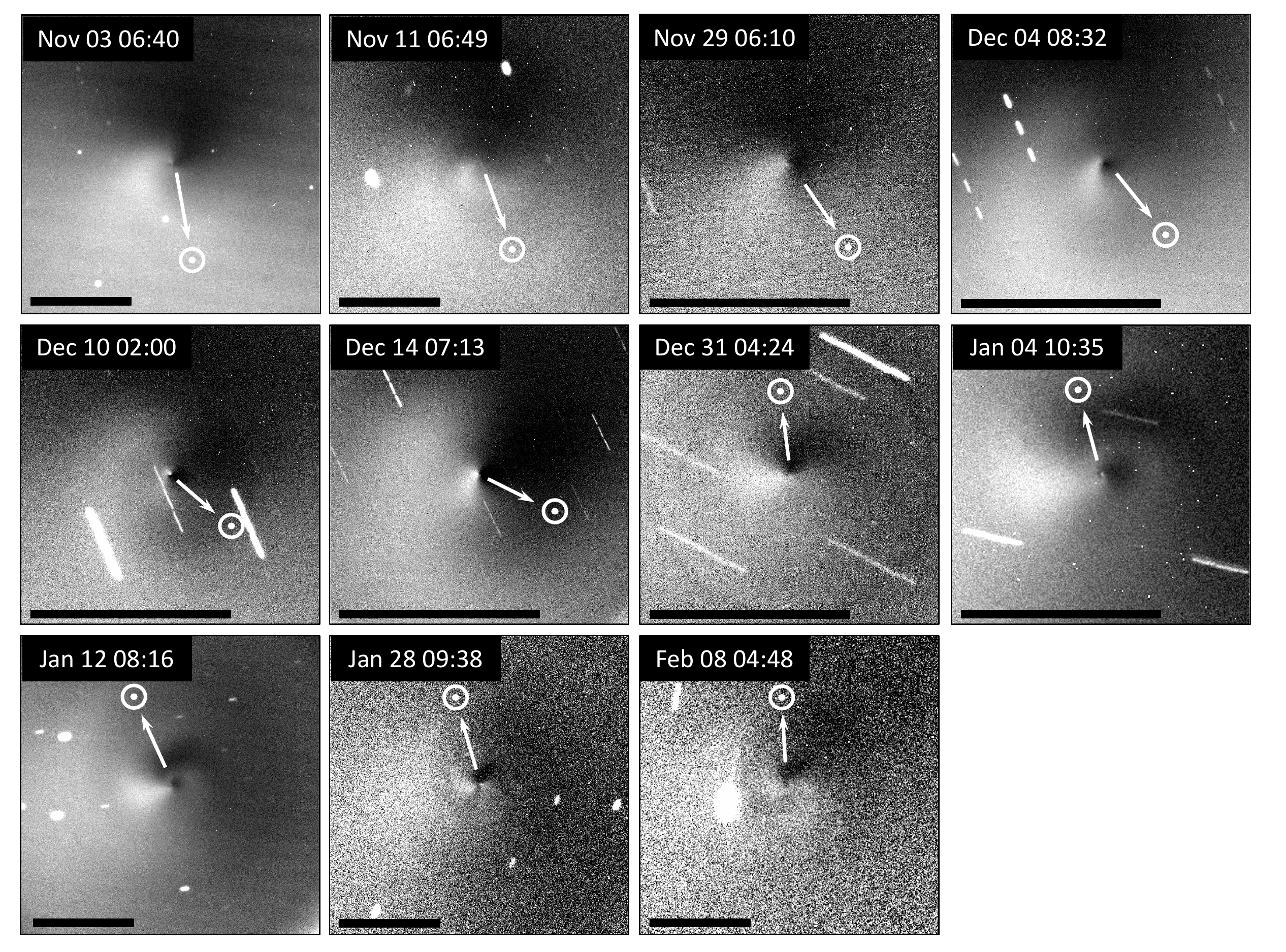}
\caption{Sample image from each observing run, showing the typical
  progression of the morphology throughout the apparition (e.g., the spiral
  starts in the clockwise direction, exhibits a broad corkscrew at closest approach,
  and ends in the counterclockwise direction). Images are enhanced by dividing 
  out an azimuthally averaged profile. North is up, East to the left, and the
  scale bar at the bottom is 20,000~km.
    \label{fig:sample_im}} 
\end{figure}

Sample CN images from each observing run are shown in
Figure~\ref{fig:sample_im}.  Throughout our observations, the basic CN
morphology is indicative of a nucleus with at least two isolated active
areas.  In general, one feature appears to have been active (at varying
levels) throughout most of a rotation, while the second turned on and off
with rotation.  As viewed from Earth during the comet's approach and
recession, the jets produced spirals around the nucleus.  The sense of
rotation was clockwise pre-perihelion and counterclockwise post-perihelion,
indicating that the Earth crossed the comet's equator sometime around
perihelion.  In the weeks around perihelion, the structures from the two
sources were broader (due to the small spatial scale caused by proximity to
Earth) and often overlapped, confusing the interpretation of the morphology.
At one point in the rotational phase, however, a corkscrew morphology is
apparent, indicating that one of the jets was at a mid-level latitude, with
the Earth outside the cone being swept out by that jet.  At other phases,
(e.g., the December~14 image in Figure~\ref{fig:sample_im}) symmetric
features are seen on opposite sides of the nucleus, suggesting that the other
active region was near the equator, with its jet sweeping across the line of
sight.  The discontinuity introduced by the overlapping features around
perihelion interferes with our ability to interpret the comet's overall
behavior.  It is not clear from inspection of the data alone, whether the two
jets seen pre-perihelion were the same as those seen post-perihelion, or if
there were more than two active areas, with different sources turning off/on
during the period of confused morphology.  See
  \cite{KnightEtal:wirtanen} for a more detailed depiction and analysis of
  the jet morphology.

The radial distance of the arcs as a function of time/rotational phase
reflects the projected velocity of the CN streaming outward from the active
areas.  In images from November~11-13 and January~12, we measured the radial
distance of the arcs as a function of time at eight position angles (PAs)
around the nucleus, and fit a linear function to each PA to estimate the gas
velocities.  In November, the highest velocity (presumably that with the
smallest projection effect) was 0.62~km~s$^{-1}$ at PA~180\degr, and in
January, the highest velocity was 0.80~km~s$^{-1}$ at a PA~90\degr.  We
also attempted these measurements with the mid-December runs, but our
attempts to consistently define radially expanding features were difficult,
due to the broad and overlapping morphology.  In this case, we do not believe
we obtained any measurements from which velocities can be reliably computed.

\subsection{Rotational Phasing and Period Determination  
\label{per_det}}

We used the comet's coma morphology as a tool for deriving the nucleus'
rotation period.  This process assumes that the nucleus was in or near a
state of simple rotation and that active areas producing the jets reacted to
the solar irradiation in the same manner on every rotation.  Thus, when a
pair of images show the same morphology, it indicates that an integer number
of rotations have passed between the two images.  In practice, the morphology
can be affected by changing illumination conditions, as the comet orbits the
Sun, and changing viewing geometry, as the comet passes the Earth. Typically
these changes are gradual and can be neglected for observations obtained over
the course of a week or two, though during Wirtanen's close approach to
Earth, they become more pronounced.  Because we have multiple observing runs
between November 2018 and February 2019, we were able to derive independent
rotation periods for different times throughout this window and use them to
look for an evolution in the rotation state as the comet passed perihelion.
For investigating the rotational phasing at different epochs, we
combined our data into groups by 
individual observing runs (denoted by numbers in the Phase Group
column in Table~\ref{tab:obs}) and by neighboring inter-run groups, when the
geometry does not change dramatically (denoted by letters).

\begin{figure}[hbt!]
\epsscale{0.5}
\plotone{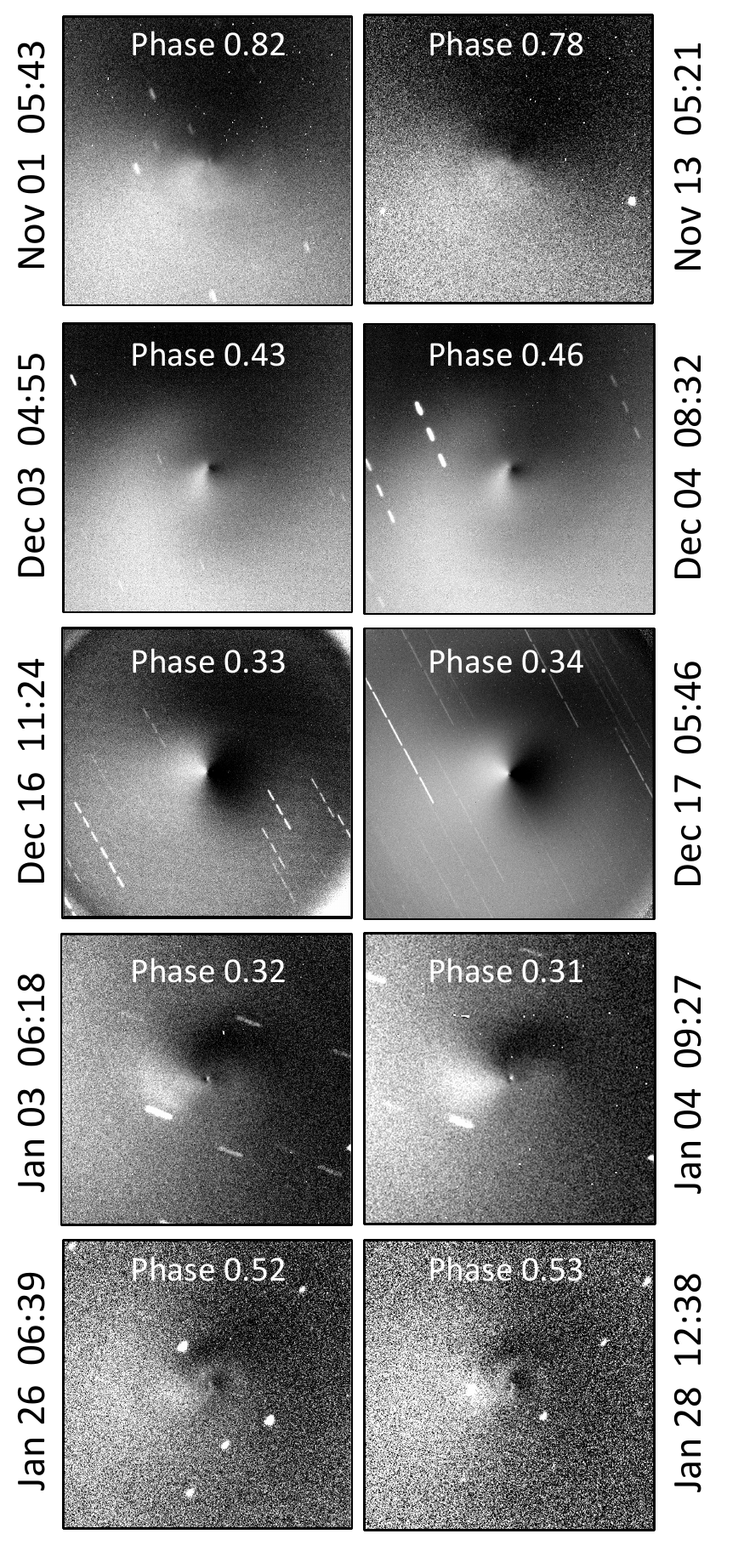}
\caption{Representative pairs of CN images used to derive the rotation
  period. The close match between the features indicates an integer number of
  rotations in the interval.  Images were enhanced by dividing out an
  azimuthally averaged profile. The phase designations are assigned, as
  discussed in the text, for reference.   
  \label{fig:im_pairs}}  
\end{figure}

We used two techniques to derive the period from the morphology, with two
authors independently taking different approaches that resulted in consistent
results.  First, we know the rotation is always near 9~hr from previous work
\citep[e.g.,][]{Farnham:cbet4571,JehinEtal:cbet4585} and from numerous nights
where we have $>$9~hr of coverage.  With this constraint, one author searched
by eye for one or more pairs of images from each group for which
distinctive morphological features matched, and derived a period by assuming
an integer number of rotations over the intervening time.  Examples of these
pairs are shown in Figure~\ref{fig:im_pairs}. This technique was used to
provide initial working periods before results from the other, more
rigorous technique was finalized.  The accuracy from this pairwise
fitting is dependent on how close in phase the image pairs end up, though
the wealth of images allowed us to find matches close enough that the 
results agreed well with our other techniques. 

In our second method, another author incorporated all of the data from a
given run, assuming a period, computing the rotational phase for each image
in that run, and then assembling an animated sequence with the images ordered
by their respective phases. Zero phase is always defined at the comet's
perihelion date, 2018 December 12.931, so the phasing for a given run will
change depending on the rotation period, and a particular phase for one run
will not match that phase in runs with different periods.  When the assumed
period matches the actual period, a movie of this type should produce a
smooth and continuous sequence of motion as the jet material flows outward
from the nucleus.  On the other hand, out-of-sequence frames (features
jumping forward and back) indicate that the assumed period is not correct,
and the number and size of these jumps grows as the difference between the
assumed period and the actual period increases.  We stepped through potential
periods at intervals of 0.01~hr to look for acceptable sequences, which
allowed us to define the range of valid periods for each run.  Because the
various enhancements reveal different aspects of the coma, we produced
animations for all three of our techniques to confirm that they give a
consistent result (these sequences can be found in animated GIF format
  at the University of Maryland (UMD) Digital
  Repository\footnote{https://drum.lib.umd.edu/handle/1903/26472}). Because
this technique uses much more data than the pairwise matching, it provides a
more precise result.

We recognize that there is an inherent subjectivity to defining an acceptable
solution, especially in sequences where variable data quality, insufficient
sampling rates, or changes in the viewing geometry can affect the apparent
timing of a feature's repetition.  Thus, to avoid rejecting potentially valid
periods, we used a conservative definition of ``smooth'' feature motions,
pushing our solutions into the range where they may exhibit a few
discontinuities to allow for these issues. These constraints are especially
relaxed around the time of close approach, where the viewpoint was changing
by as much as 4\degr\ per day.  To minimize the effects of rapidly changing
geometry for our December 13-17 observations, we separated the data into
stepwise groups of 3 nights, deriving a separate period for each group.
Although we obtained data on December 12 as well, these images were not
included in this grouping due to the interference of an outburst.  The
division between acceptable and unacceptable values typically occurs when a
period increment causes one or more pairs of key images to flip their
sequence order, revealing an obvious discontinuity in the motion that cannot
possibly be attributed to observing conditions or geometry changes.  We
define the measurement's uncertainty as the point between the marginally
acceptable value and the obviously unacceptable value (which effectively
means our uncertainties are at the 3-$\sigma$ level).  Thus, the center of
our range of periods represents the smoothest sequence of images (our best
estimate of the apparent period) while the uncertainties encompass any values
that could be valid given the natural complexities of the data.

Finally, we combined data from the inter-run groups to refine the period even
further. These groupings proved very powerful for several reasons.  First,
they increase the number of images used in each sequence, improving the phase
coverage and overlap.  Second, they extend the time baseline from a few days
to a week or two, which, because small changes in period are amplified by the
large number of intervening cycles, improves the precision.  Finally, we have
four observing runs in which we were unable to derive periods due to an
insufficient number of images for reliable phasing, and combining each of
these runs with a neighboring run allows us to incorporate these data.  (Even
a few images, when interleaved with a more complete sequence, can be very
constraining.)  For the early November and late January time frames, the
geometry changes are minimal and thus these runs can be reliably combined.
In early and late December, the changing geometry between runs, although not
extreme, becomes noticeable, so we accepted a wider range of periods to
account for these potential effects.  Even so, the results from these
inter-run combinations tend to be consistent with the individual runs from
which they are comprised, but with smaller uncertainties.

\begin{figure}[hbt!]
\epsscale{0.75}
\plotone{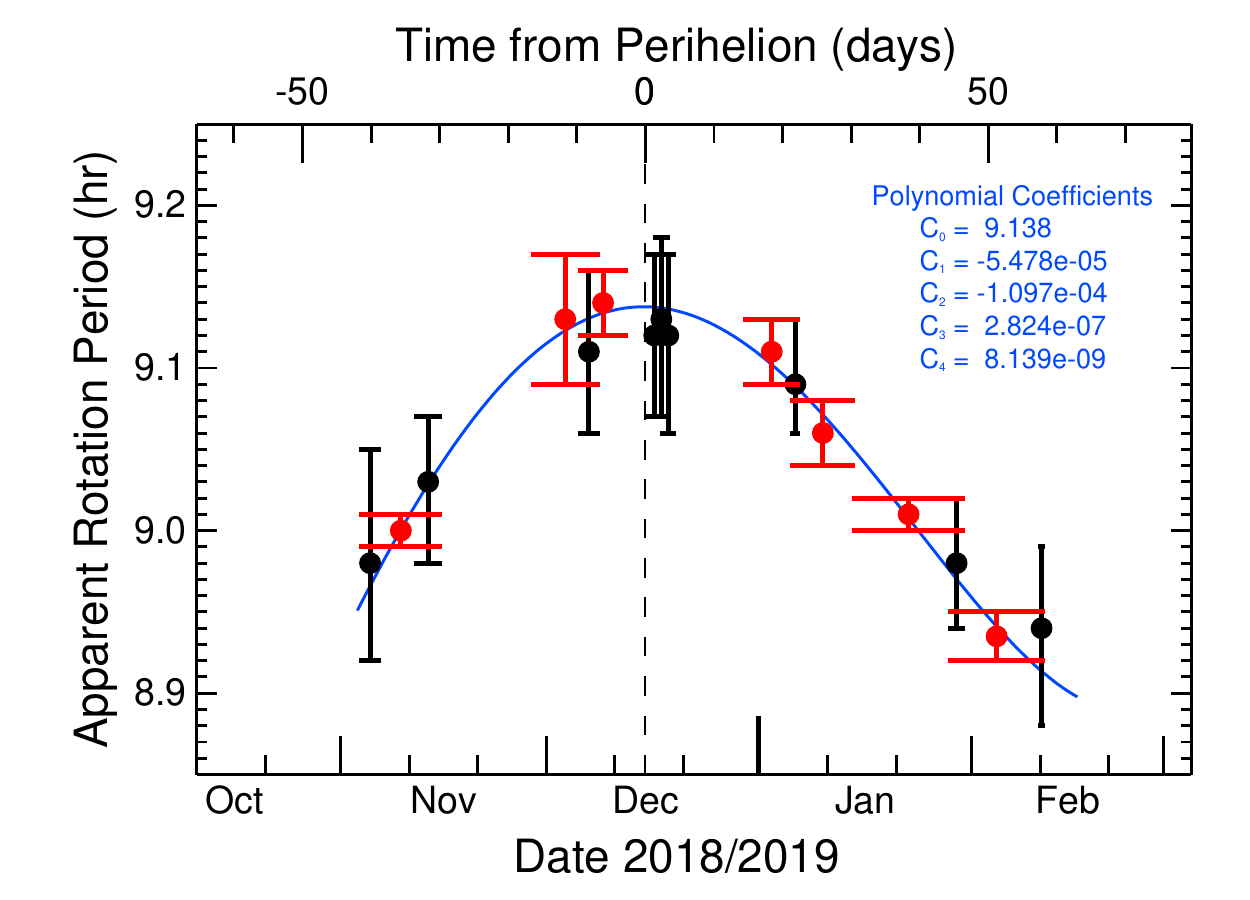}
\caption{Rotation periods as a function of time derived from the image
  sequences. The vertical bar on each point indicates the range of acceptable
  solutions and the horizontal crossbars indicate the range of dates used for
  that sequence. Black symbols are solutions from individual runs, while the
  red symbols represent inter-run combination results. The blue curve shows
  a 4th order polynomial fit to the data. 
  \label{fig:periods}}
\end{figure}

\begin{figure}[hbt!]
\epsscale{1.15}
\plotone{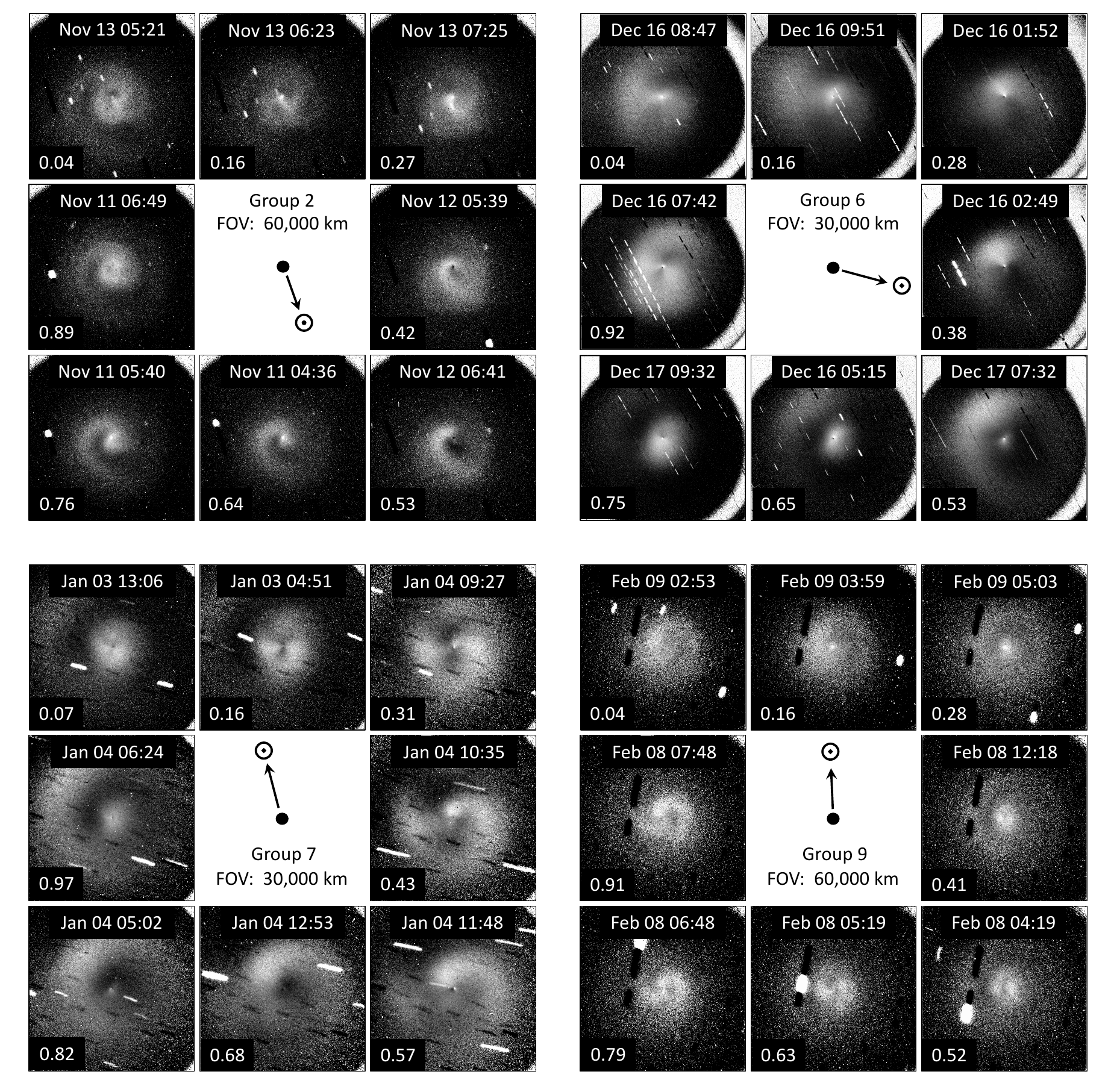}
\caption{Representative looped image sequences showing the changing CN
  morphology as a function of rotational phase.  Images were enhanced using
  the temporally averaged mask and phased to the polynomial fit values for
  each group (Table~\ref{tab:periods}). North is up and East to the
  left. Full sequences for all groups of data can be found in animated GIF
  format at the UMD Digital repository https://drum.lib.umd.edu/handle/1903/26472. 
  \label{fig:sequences}}
\end{figure}

The periods we derived are listed in Table~\ref{tab:periods} and plotted in
Figure~\ref{fig:periods}.  These results suggest that the apparent period
increased by $\sim$0.15~hr in the five weeks before perihelion, peaking at
9.14~hr before decreasing again by $\sim$0.2~hr in the eight weeks after
perihelion.  A 4th-order polynomial was fit to the measurements
(coefficients: [9.138, -5.478$\times10^{-5}$, -1.097$\times10^{-4}$, 2.824$\times10^{-7}$, 8.139$\times10^{-9}$]) to provide
a continuous representation of the period as a function of time, and we adopt
the values from this curve to provide consistency in the different
presentations of phased data throughout this paper.  Figure
Set~\ref{fig:sequences} shows sequences of images from the individual
observing runs, phased to the polynomial fit period for each run.  Overall,
the coma morphology repeats consistently over the course of single and
multiple rotations (within the constraints introduced by the geometry
changes) strongly suggesting that there are no noticeable effects produced by
non-principal axis rotation.  Thus, we conclude that the nucleus is in a
state of simple rotation or nearly so.  

\begin{deluxetable*}{clrrcccc}[hbt!]
\tablecaption{Rotation Periods Derived from Morphology \tablenotemark{a} \label{tab:periods}}
\tablecolumns{8}
\tablewidth{0pt}
\tablehead{
\colhead{Phase} &
\colhead{Date Range} &
\colhead{$\Delta t_{\rm start}$\tablenotemark{c}} &
\colhead{$\Delta t_{\rm end}$\tablenotemark{c}} &
\multicolumn{3}{c}{Period (hr)\tablenotemark{d}} & 
\colhead{Poly.\tablenotemark{e}} \\ \cline{5-7}
\colhead{Group\tablenotemark{b}} & \colhead{} & \colhead{(day)} &
\colhead{(day)} & \colhead{Min} & \colhead{Best} & \colhead{Max} &
\colhead{(hr)}}
\startdata
1  & 2018 Nov 01-04                      &      --41.69  &      --38.60  &      8.92  &      8.98  &      9.05  &      8.97  \\
{\bf A}  & {\bf 2018 Nov 01-13}          & {\bf --41.69} & {\bf --29.62} & {\bf 8.99} & {\bf 9.00} & {\bf 9.01} & {\bf 9.00} \\
2  & 2018 Nov 09-13                      &      --33.72  &      --29.62  &      8.98  &      9.03  &      9.07  &      9.03  \\
---& 2018 Nov 26-29                      &      --16.73  &      --13.67  &       ---  &       ---  &       ---  &      9.11  \\
{\bf B}  & {\bf 2018 Nov 26-Dec 06}      & {\bf --16.73} &  {\bf --6.65} & {\bf 9.09} & {\bf 9.13} & {\bf 9.17} & {\bf 9.12} \\
3  & 2018 Dec 03-06                      &       --9.82  &       --6.65  &      9.06  &      9.11  &      9.16  &      9.13  \\
{\bf C}  & {\bf 2018 Dec 03-10}          &  {\bf --9.82} &  {\bf --2.52} & {\bf 9.12} & {\bf 9.14} & {\bf 9.16} & {\bf 9.13} \\
---& 2018 Dec 09-10                      &       --3.85  &       --2.52  &       ---  &       ---  &       ---  &      9.14  \\
4  & 2018 Dec 13-15                      &         0.13  &         2.46  &      9.07  &      9.12  &      9.17  &      9.14  \\
5  & 2018 Dec 14-16                      &         1.12  &         3.54  &      9.07  &      9.13  &      9.18  &      9.14  \\
6  & 2018 Dec 15-17                      &         2.15  &         4.54  &      9.06  &      9.12  &      9.17  &      9.14  \\
---& 2018 Dec 27-31                      &        14.23  &        18.27  &       ---  &       ---  &       ---  &      9.11  \\
{\bf D}  & {\bf 2018 Dec 27-2019 Jan 04} &  {\bf  14.23} &  {\bf  22.62} & {\bf 9.09} & {\bf 9.11} & {\bf 9.13} & {\bf 9.10} \\
7  & 2019 Jan 03-04                      &        21.15  &        22.62  &      9.06  &      9.09  &      9.13  &      9.09  \\
{\bf E}  & {\bf 2019 Jan 03-12}          &  {\bf  21.15} &  {\bf  30.61} & {\bf 9.04} & {\bf 9.06} & {\bf 9.08} & {\bf 9.07} \\
---& 2019 Jan 12                         &        30.15  &        30.61  &       ---  &       ---  &       ---  &      9.05  \\
{\bf F}  & {\bf 2019 Jan 12-28}          &  {\bf  30.15} &  {\bf  46.61} & {\bf 9.00} & {\bf 9.01} & {\bf 9.02} & {\bf 9.01} \\
8  & 2019 Jan 26-28                      &        44.16  &        46.61  &      8.94  &      8.98  &      9.02  &      8.97  \\
{\bf G}  & {\bf 2019 Jan 26-Feb 09}      &  {\bf  44.16} &  {\bf  58.29} & {\bf 8.92} & {\bf 8.935}& {\bf 8.95} & {\bf 8.94} \\
9  & 2019 Feb 08-09                      &        57.25  &        58.29  &      8.88  &      8.94  &      8.99  &      8.91  \\
\enddata
\tablenotetext{a}{ Periods determined from each observing run, and in boldface, from inter-run combinations. Nov 26--29, Dec 09-10 and Jan 12 did not have sufficient data to produce reliable period measurements, but were used in inter-run combinations.}
\tablenotetext{b}{ Morphology Phase Group listed in Table~\ref{tab:obs} used to combine data for phasing.}
\tablenotetext{c}{ Start and end times, relative to perihelion, of the data in the group}
\tablenotetext{d}{ Minimum, best, and maximum periods that produce acceptable sequences}
\tablenotetext{e}{ Period derived from the 4th order polynomial fit}
\end{deluxetable*}

\subsection{Exploration of Synodic Effects \label{sec:synodic}}

The fact that the changes in the apparent rotation periods are symmetric
around the time of close approach raises the question of whether these
changes are real or are they caused by synodic effects from the rapidly
changing viewpoint?  We explored this question both qualitatively and
quantitatively.  Some synodic effects arise when changes in the viewing
geometry either shorten or lengthen the time for a reference longitude to
return to the same spot relative to the observer.  To evaluate the
contributions from synodic effects, we look at the extremes where the
geometry remained nearly constant or where it changed rapidly.  In early
November and late January/February, the comet's motion was primarily
toward/away from the Earth, with minimal change in viewpoint, so the apparent
period from these times should be close to the sidereal period
($<$0.001~hr/rot).  However, our measured periods are changing most rapidly
during these epochs, suggesting that the sidereal period itself is evolving.
In addition, our November measurement is different from that seen in
February, which also argues that the sidereal period has changed through
perihelion.  At the other extreme, the biggest synodic effects should have
occurred in the weeks surrounding closest approach when the viewing geometry
changed most rapidly ($\sim$4\degr/day), yet our measurements show fairly
constant values during this time frame.  This contradicts the idea that the
variations are produced by the viewing geometry.

\begin{figure}[hbt!]
\epsscale{0.75}
\plotone{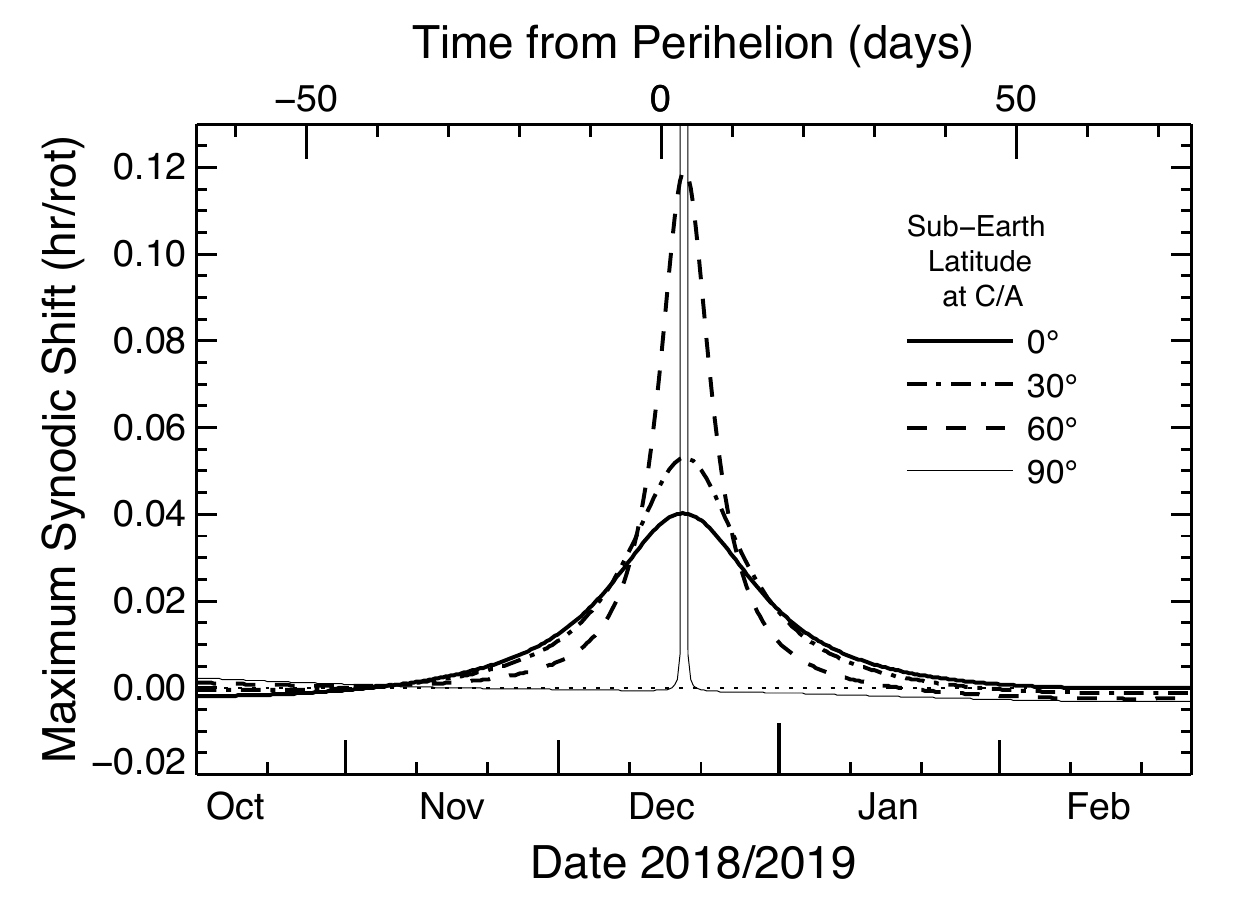}
\caption{Maximum possible shifts between Wirtanen's apparent and sidereal
  periods due to the changing viewing geometry from the Earth.  Synodic
  effects are plotted for four different pole orientations: one with the Earth
  at the equator (0\degr) and three that allow the Earth to reach
  30\degr, 60\degr, and 90\degr\ latitude (see Section~\ref{sec:synodic} for
  additional discussion). These results, when compared the results shown
  in Figure~\ref{fig:periods}, show that the synodic effects are too
  small in magnitude and too narrow temporally to have produced the observed
  period changes.
  \label{fig:synodic}}
\end{figure}

We also performed more rigorous calculations to explore the viability of
geometric effects producing our apparent measurements.  The maximum possible
synodic effects will occur if the fastest relative motion (e.g., at closest
approach) corresponds to the time when the observer is at its highest
latitude (where the cos(latitude) term magnifies the longitudinal motion).
Because the sub-Earth latitude is determined by the spin axis orientation, we
performed calculations for four different pole positions, one where the Earth
skims along the comet's equator, and three in which the sub-Earth point peaks
at latitudes of 30\degr, 60\degr, and 90\degr.  (Synodic effects for pole
orientations in the opposite direction shorten the rotation period, which is
the reverse of the trend we observed.)  For each case, the synodic effects as
a function of time are plotted in Figure~\ref{fig:synodic} showing a
trade-off between their magnitude and the duration that they act (i.e., as
the peak increases, the curve gets narrower).  Comparing these calculations
to the results in Figure~\ref{fig:periods} shows that the changes seen in our
measured periods cannot be due exclusively to the changing geometry.  Not
only are the magnitudes of the synodic effects too small (for all but the
highest latitude cases) but their contributions are limited to a window
around close approach that is too short to explain the trends that we see
(confirming our qualitative analysis).  We explored the synodic effects
induced by the Sun for the same pole orientations, but these are
substantially smaller than the Earth's effects (peaking at
$\sim$0.01~hr/rot), and so they too, are insufficient for explaining the
observed period variations. Thus, we conclude that the changes seen in our
measurements were actual changes in the nucleus' rotation period.

\section{CN Lightcurves and Rotational Analysis}  \label{sec:lightcurves}

Wirtanen's lightcurves offered a second method for measuring the comet's
rotation period.  Although somewhat hampered by calibration issues, as
discussed below, this technique gives a separate measure of the spin period,
providing a check on the morphology results.  Furthermore, because the
photometry is less dependent on the Earth's motion, only the low-level
solar synodic effects will apply, producing a result closer to the sidereal
value at close approach.

Although we have good temporal coverage on many nights with both R (or {\em
  r\,$^\prime$}) and CN filters, we chose to use the CN images because they
consistently showed evidence of rotational variability (due to the higher gas
velocities that allow the CN to leave the measuring aperture more rapidly,
enhancing the amplitude of the variations).  Because of weather and geometric
circumstances, we had relatively few nights that could be fully calibrated
(e.g., even on clear nights, the proximity to Earth means that, in much of
our data, the coma fills the field of view, precluding an accurate measure
of the sky background).  Thus, for our period determinations, we decided to
forego an absolute calibration and focus on the relative brightness changes
and the timing of the peaks and troughs within a night.  For this reason, we
did not restrict our sample to photometric conditions but also accepted
nights of fairly good quality that could be corrected, as discussed in
section~\ref{sec:photcal}.  The nights used in our lightcurve analyses are
listed, with nightly conditions, in Table~\ref{tab:phot_obs}.

\begin{deluxetable*}{lccccc}
\renewcommand{\baselinestretch}{0.9}
\tablecaption{Photometry Observing Conditions\tablenotemark{a} \label{tab:phot_obs}}
\tablecolumns{6}
\tablewidth{0pt}
\tablehead{
\colhead{Date} &
\colhead{Avg. Bright.} &
\colhead{$\Delta m$\tablenotemark{c}} &
\colhead{$\sigma_m$\tablenotemark{d}} &
\colhead{Seeing\tablenotemark{e}} &
\colhead{Phase} \\ 
\colhead{} & \colhead{(mag)\tablenotemark{b}} & \colhead{(mag)} & 
\colhead{(mag)} & \colhead{(arcsec)} &  \colhead{Groups\tablenotemark{f}} }
\startdata
2018 Nov 11 & 12.506 &   ---   & 0.002 & 3.0 &   2   \\
2018 Nov 12 & 12.447 &  0.010  & 0.002 & 4.4 &   2   \\
2018 Nov 13 & 12.373 &   ---   & 0.002 & 4.1 &   2   \\
2018 Dec 12 & 11.738 &  0.060  & 0.001 & 1.3 &   5   \\
2018 Dec 13 & 11.762 &  0.025  & 0.001 & 2.7 &   5   \\
2018 Dec 14 & 11.798 &   ---   & 0.001 & 1.5 &   5   \\
2018 Dec 16 & 11.760 &  0.035  & 0.001 & 1.2 &   5   \\
2018 Dec 17 & 11.809 &  0.075  & 0.001 & 1.8 &   5   \\
2019 Jan 03 & 11.866 &   ---   & 0.002 & 4.4 &  7,E  \\
2019 Jan 04 & 11.864 &   ---   & 0.002 & 2.8 &  7,E  \\
2019 Jan 12 & 12.440 &  0.040  & 0.001 & 2.2 &  E,F  \\
2019 Jan 26 & 12.982 &  0.015  & 0.003 & 3.3 & 8,F,G \\
2019 Jan 28 & 12.997 & -0.005  & 0.003 & 2.2 & 8,F,G \\
2019 Feb 08 & 13.630 &   ---   & 0.003 & 4.4 &  9,G  \\
2019 Feb 09 & 13.708 &  0.030  & 0.003 & 3.3 &  9,G  
\enddata
\tablenotetext{a}{ Additional information is contained in Table~\ref{tab:obs}}
\tablenotetext{b}{ Average brightness of the lightcurve for each night, used for aligning
  different nights.}
\tablenotetext{c}{ Magnitude offset applied to align lightcurves from different nights.}
\tablenotetext{d}{ Typical photometric uncertainty for the night.}
\tablenotetext{e}{ Typical FWHM seeing for the night.}
\tablenotetext{f}{ Groups used to phase data in the photometry
  analyses; Numbers link nights combined over a single observing run,
  Letters combine nights over two runs. Selected to match the groups in
  Table~\ref{tab:periods} to facilitate comparisons, though because geometry
  is less of an issue for photometry, we combine all the mid-December data
  into a single group, 5, that can be compared to groups 4-6 in the morphology.}
\end{deluxetable*}
\renewcommand{\baselinestretch}{1.0}


\subsection{Photometric Calibrations} \label{sec:photcal}

To assemble our lightcurves, we started with the bias-removed and
flat-fielded images, centroided on the optocenter of each image and performed
photometry using a 10-arcsec radius aperture.  This aperture size was used
throughout the apparition as a compromise between the need for an aperture
large enough to minimize the effects of seeing variations, but small enough
to enhance the rotational variability. (Because of the large range in
$\Delta$, an aperture of fixed physical size at the comet was impractical.)
This results in a different fraction of the coma being measured on each
night, but as our objective is the variability within the night, this is a
minor concern.

The sky background level was estimated using an annulus centered on the
optocenter.  The outer radius is set independently on each night, using the
maximum dimension that can be used throughout that night (avoiding chip
edges, vignetting, etc.), while the width of the annulus was $\sim$100 pixels.
Coma fills the field in December (and into November and January), so the sky
level is usually over-estimated in our measurements, but because it tends to
be stable during each night, it should have a minimal effect on the
rotational variability analyses.

On photometric nights, we used measured coefficients to correct for
extinction.  The remaining nights used coefficients from the same run, if
available, or typical coefficients for the telescope as a last resort. On
runs where the sky contamination was minimal, we could also use field stars
to correct for the relative extinction due to airmass and thin cirrus
throughout a night \citep[e.g.,][]{KnightEtal:T2rotation,KnightEtal:T2study,
  SchleicherEtal:T2rotation,EisnerEtal:arend}.  Such tweaks were not possible
during December and early January, due to rapid proper motions and extreme
coma contamination, so there may be residual extinction trends on those
nights.  Fortunately, the majority of those data were acquired at
airmass $<$1.5, so the trends are unlikely to affect our interpretations.

Typically, our final task was the removal of underlying continuum from the
calibrated CN images \cite{FarnhamEtal:hbfilters}. However, as noted earlier,
the continuum signal in most of our observations was minimal, and because we
were unable to remove the continuum from all of our data, we decided to stay
consistent and not remove it from any of our data.  If any signal from the
dust is detected, it would slightly dampen the amplitude of the lightcurve
variations, but should have minimal effect on the timing of the peaks and
troughs, and thus the period determinations.

\subsection{Photometry Results \label{sec:phot_res}}

\begin{figure}[hbt!]
\epsscale{1.1}
\plotone{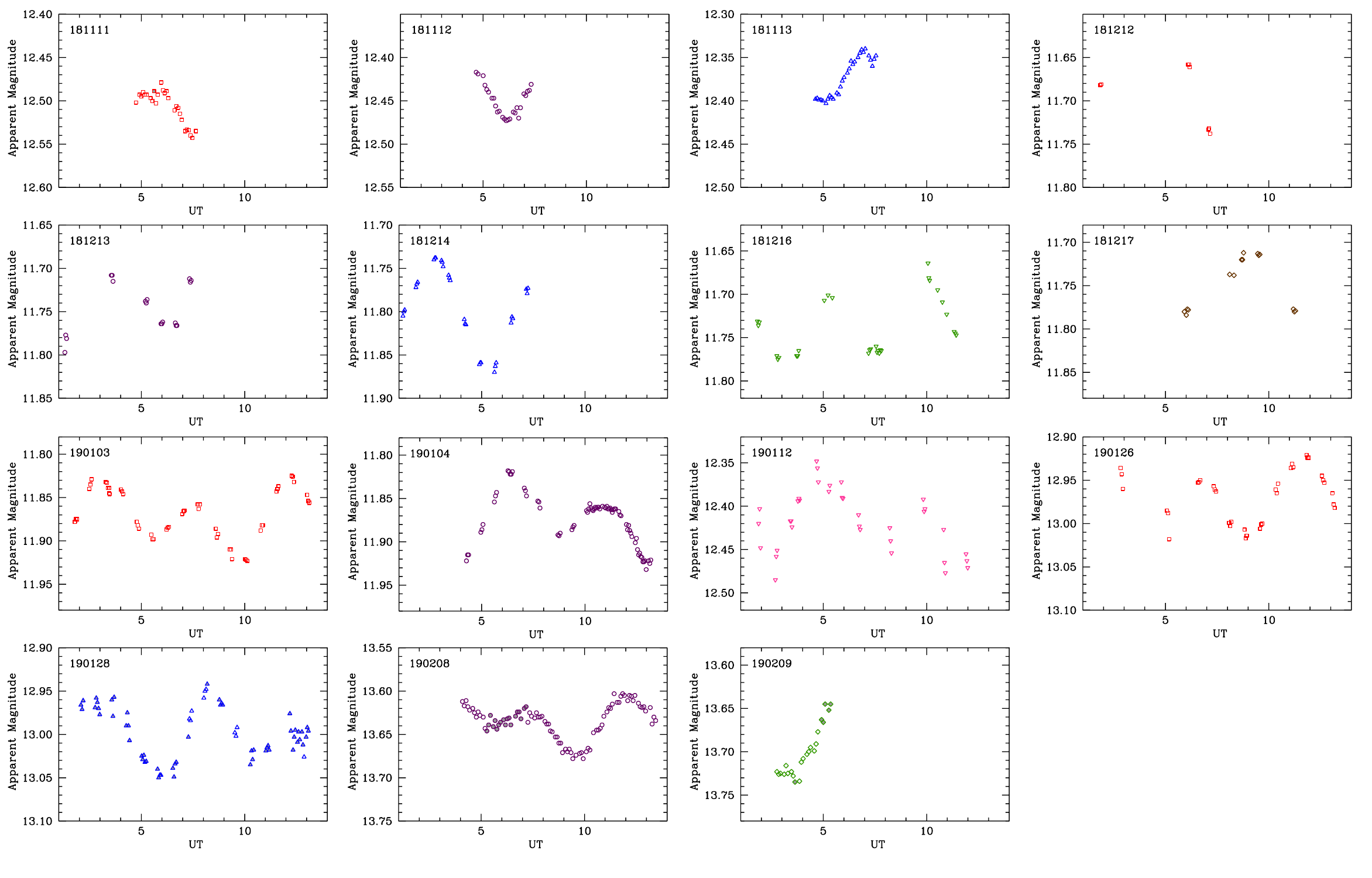}
\caption{Lightcurves segments from our nightly observations (dates listed in
  YYMMDD format).  Points filled in grey have been corrected for extinction
  using field star comparisons.  Observation times have been corrected for
  light travel time.  The CN magnitude data used to produce these plots is
  included as Data Behind the Figure.
  \label{fig:nightlight}}
\end{figure}

Our nightly CN lightcurves are plotted in Figure~\ref{fig:nightlight} (the
photometry measurements are available as Data Behind the Figure).  Data points
that are filled in gray denote that an extinction correction derived from
field star extinction was applied during the calibration process.  A
number of nights (December 16, January~3 \&~4, February~8, etc.)  had
regular, high quality data spanning $\gtrsim$9~hr.  Although these nights are
valuable for confirming the $\sim$9~hr period and for permitting assessment
of the full lightcurve shape in one night's observations, they are of limited
value in deriving precise rotation periods.  For period determination, we phased
multiple nights' data (combined as noted in the Phase Groups
column of Table~\ref{tab:phot_obs}) to construct more extensive lightcurves.


\begin{figure}[hbt!]
\epsscale{0.75}
\plotone{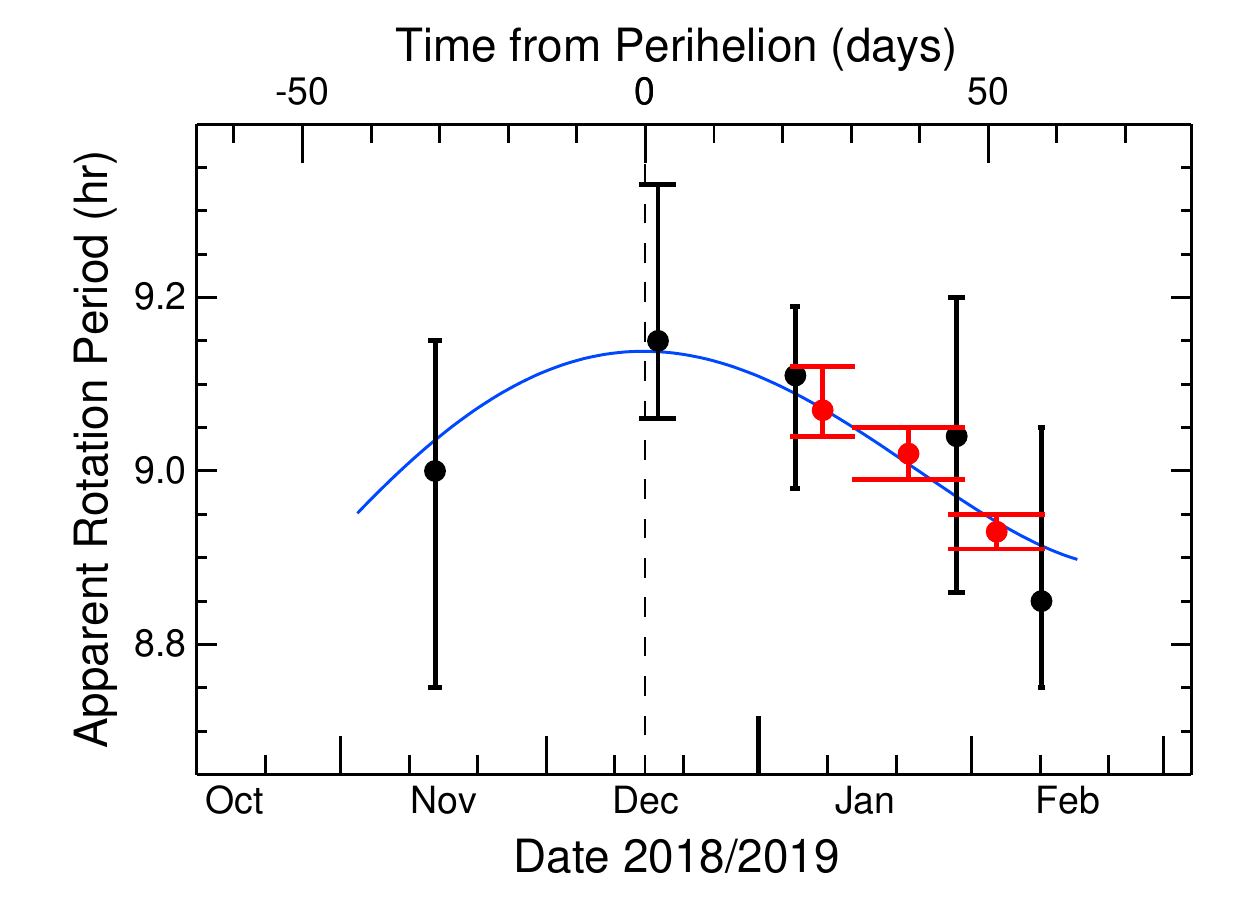}
\caption{Rotation periods as a function of time derived from the
  photometric lightcurves, showing the same trends as in the
  morphology-derived periods, but with larger uncertainties. The error
  bars and colors represent the same information stated in
  Figure~\ref{fig:periods}.  The blue curve reproduces the 4th order
  polynomial derived from the morphology sequences.   
  \label{fig:lcperiods}}
\end{figure}

We determined rotation periods by eye as discussed in
\citet{SchlKnight:LINEAR}, phasing the lightcurves to different periods and
looking for the best alignment of the overlapping segments.  While evaluating
the best fits, we allowed arbitrary vertical offsets of nightly segments
(listed in Table~\ref{tab:phot_obs}) to account for any brightness
differences arising from our lack of absolute calibration.  After the best
period was found, we estimated its uncertainties by exploring how much the
period could be changed before the lightcurves showed an obvious misalignment
(roughly a 3-$\sigma$ uncertainty).  Uncertainties are dependent on the
baseline of the observations, with a longer span of observations producing
better precision, but the shape of the lightcurve changes over time, so we
limited our measurements to groups of data spanning less than two weeks.  The
periods derived from our lightcurve analyses are listed in
Table~\ref{tab:lc_properties} and plotted in
Figure~\ref{fig:lcperiods}. These results are in excellent agreement with
those derived from our morphology analysis, but with larger uncertainties due
to the calibration issues.  This confirms that the comet's rotation rate was
changing, and also indicates that the synodic effects introduced by the
Earth's motion were minimal.


\begin{deluxetable*}{clcccccc}[hbt!]
\renewcommand{\baselinestretch}{0.9}
\tablecaption{CN Lightcurve Properties\tablenotemark{a} \label{tab:lc_properties}}
\tablecolumns{8}
\tablewidth{0pt}
\tablehead{
\colhead{Phase} &
\colhead{Nights Used} &
\colhead{Mid-Date\tablenotemark{c}} &
\colhead{Period\tablenotemark{d}} &
\colhead{P. Range\tablenotemark{e}} &
\colhead{Poly. Fit\tablenotemark{f}} &
\colhead{L. Range\tablenotemark{g}} &
\colhead{Phase} \\
\colhead{Group\tablenotemark{b}} &
\colhead{} &
\colhead{}  &
\colhead{(hr)} &
\colhead{(hr)} &
\colhead{(hr)} &
\colhead{(mag)} &
\colhead{Sep.\tablenotemark{h}}}
\startdata
2  & Nov 11, 12, 13            & Nov 12.256 & 9.00 & 8.75--9.15 & 9.03 & 0.065 & 0.50  \\
5  & Dec 12, 13, 14, 16, 17    & Dec 14.774 & 9.15 & 9.06--9.33 & 9.14 & 0.125 & 0.54  \\
7  & Jan 3, 4                  & Jan ~3.812 & 9.11 & 8.98--9.19 & 9.09 & 0.105 & 0.54 \\
{\bf E}  & {\bf Jan 3, 4, 12}  & {\bf Jan ~7.787} & {\bf 9.07} & {\bf 9.04--9.12} & {\bf 9.07} & {\bf 0.105}  & {\bf 0.54}  \\
{\bf F} & {\bf Jan 12, 26, 28} & {\bf Jan 20.312} & {\bf 9.02} & {\bf 8.99--9.05} & {\bf 9.01} & {\bf 0.095} & {\bf 0.52} \\
8  & Jan 26, 28                & Jan 27.332 & 9.04 & 8.86--9.20 & 8.97 & 0.095 & 0.52  \\
{\bf G} & {\bf Jan 26, 28, Feb 8, 9} & {\bf Feb ~2.171} & {\bf 8.93} & {\bf
  8.91--8.95} & {\bf 8.94} & {\bf 0.090} & {\bf 0.50} \\
9  & Feb 8, 9                  & Feb ~8.697 & 8.85 & 8.75--9.05 & 8.91 & 0.075 & 0.50  \\
\enddata
\tablenotetext{a}{ Periods and lightcurve properties determined from each observing run, and in boldface, from inter-run combinations.}
\tablenotetext{b}{ Photometric Phase Groups listed in
  Table~\ref{tab:phot_obs} used to combine the lightcurves. Selected to match the groups in
  Table~\ref{tab:periods} to facilitate comparisons.} 
\tablenotetext{c}{ Date defining the midtime of the lightcurve group.}
\tablenotetext{d}{ Rotation period derived from photometry.}
\tablenotetext{e}{ Range of acceptable periods.}
\tablenotetext{f}{ Polynomial fit from the morphology measurements (adopted for
  plotting results).}
\tablenotetext{g}{ Peak-to-trough range of lightcurve brightness.}
\tablenotetext{h}{ Phase separation between primary and secondary peaks.}
\end{deluxetable*}
\renewcommand{\baselinestretch}{1.0}

\begin{figure}[hbt!]
\epsscale{0.8}
\plotone{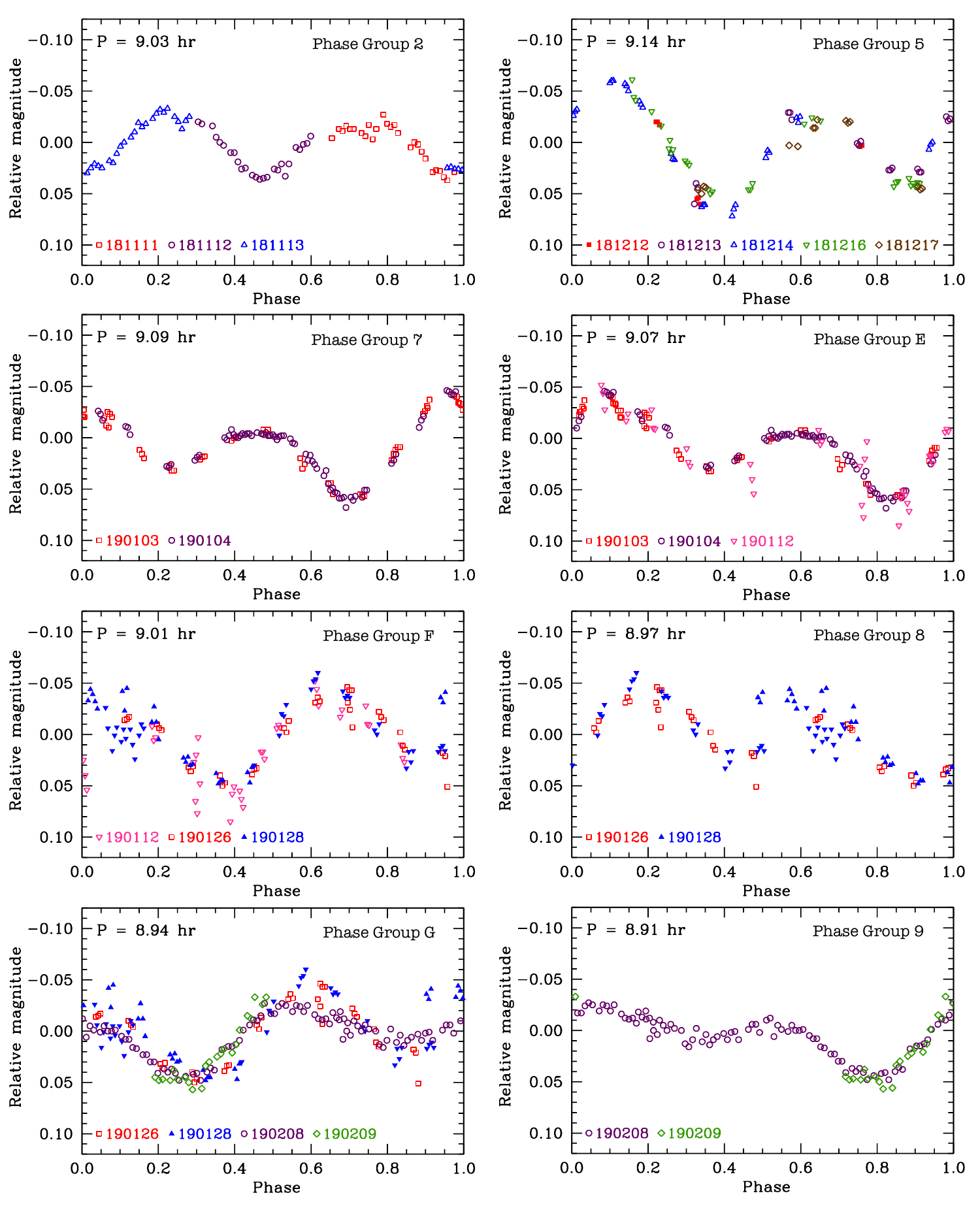}
\caption{Multi-night phased lightcurves showing the comet's
  variability. Symbols match those used in Figure~\ref{fig:nightlight},
  though in this plot, filled points indicate data that were obtained during
  outbursts.  For January~28, triangles pointing upward denote measurements
  from the first four hours of the night, while those pointing downward 
  designate the rest of the night. Periods from the 4th order polynomial fit
  (listed in each panel) were used for the phasing.  Because the different
  periods are all referenced to our fiducial point at perihelion, a given
  phase in one panel does not match that in other panels.  
  \label{fig:lightcurves}}
\end{figure}

Our multi-night phased lightcurves are shown in Figure~\ref{fig:lightcurves}.
Because the derived periods are consistent between our techniques, we have
adopted the values from our 4th order polynomial for displaying our photometry
plots as well, which allows us to compare the relative phases in the
lightcurves to the morphology from the same time.  (The results show little
difference from those using the periods derived from the lightcurve
analysis.)  These plots reveal a double-peaked lightcurve, with the
variations produced by cometary activity.  Although the shape and the
peak-to-peak range vary throughout the apparition, one peak is consistently
shallower than the other.  Table~\ref{tab:lc_properties} lists the lightcurve
ranges, which vary from $\sim$0.065--0.125~mag and the phase separation from
the higher to the lower peak.  These separations are not equidistant, but
vary between 0.50 and 0.54, with the biggest separations correlated with
close approach.

\section{Outburst Events \label{sec:outbursts}}

\begin{figure}[hbt!]
\epsscale{1.10}
\plotone{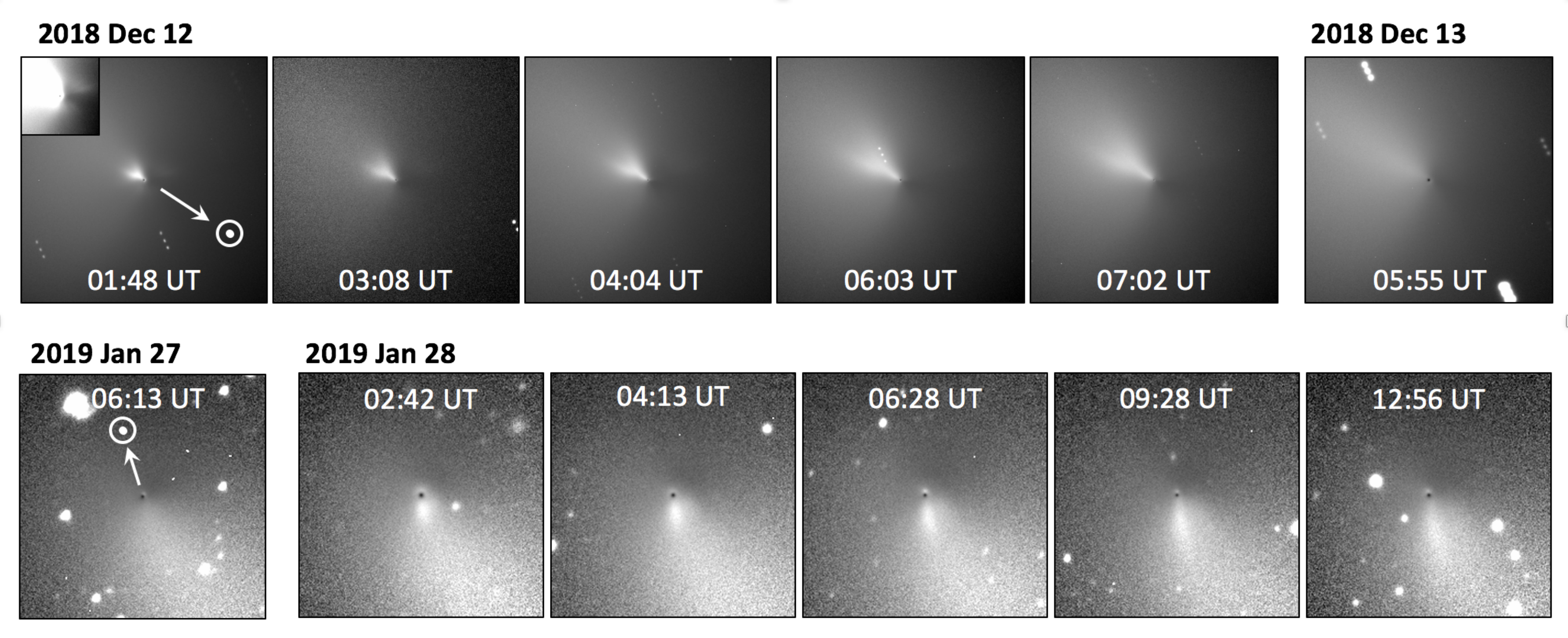}
\caption{R-band images showing the dust morphology in the December~12 and
  January~28 outbursts.  Images are enhanced by dividing out a 1/$\rho$
  profile. The December frames are 10,000~km across and the January frames 
  are 30,000~km across.  The inset in the first panel shows the inner 3,000~km
  of coma with a harsher stretch to reveal the faint ejecta to the Northwest
  and South.  The December~13 and January~27 images show the closest
  outburst-free coma for comparison.
  \label{fig:outbursts}}
\end{figure}

Our observations include two significant outbursts, one on December~12 and a
second on January~28.  The changing dust morphologies, seen in broadband R filter
sequences, are shown in Figure~\ref{fig:outbursts}.

The first outburst was dominated by a bright, ``V''-shaped extension to the
Northeast.  There is also material enveloping the Northwest and South, though
it is fainter and fades more rapidly.  Although one arm of the ``V'' was in
the anti-sunward direction, the other was not, and as there is no curvature
toward the tail, we conclude that radiation pressure was not a significant
issue at the observed distances.  Furthermore, if we assume the outburst was
in sunlight at the time, then the small solar phase angle suggests that there
are likely to be notable projection effects toward the Earth.  This is
supported by the fact that ejecta were detected at azimuths almost entirely
around the nucleus.  The linear nature of the arms of the ``V'' also suggests
that there was little effect from rotation and thus the event was of
short-duration.

We measured the motion of the outer apex of the ``V'' and derived a
projected expansion velocity 68$\pm$5~m~s$^{-1}$.  Extrapolating this
measurement back to the nucleus indicates the outburst began December~12 at
00:13~UT~$\pm$7~min.  Because we see outburst material extending into the
optocenter throughout our December~12 observations, we can constrain the
slowest moving material to speeds $<$20~m~s$^{-1}$ (projected, and assuming
an impulsive outburst).  The dust ejected in the ``V'' feature was bright enough
to dominate the signal, even in the CN filter, throughout the rest of the
night.  There is no obvious sign of the outburst in the morphology on
December~13, though given the derived speeds, diffuse residual material is
likely to be present.

The second outburst, on January~28, exhibited a narrow stream of material
flowing to the South with a slight curvature toward the anti-solar direction.
The curvature cannot be the result of rotation (which would imply an extended
period of emission), because it is opposite the direction of the spirals seen
in the CN features.  This suggests that the ejecta is composed of small dust
grains that are rapidly being pushed tailward by solar radiation pressure.
The leading edge of the material has a projected velocity of
162$\pm$15~m~s$^{-1}$, which indicates the event began January~27 at
20:01~UT~$\pm$30~min.  This outburst is notably fainter than the December
event, and is seen in our R~filter images.  Although there is no indication
of the dust stream in our enhanced CN images, the first few hours of
photometry are $\sim$0.03~mag brighter than the measurements 9~hr (one
rotation) later (cf. phase 0.5 in the January~26 \&~28 plot of
Figure~\ref{fig:lightcurves}).  This suggests that the photometry is sensitive
to a secondary contribution from the outburst, possibly as diffuse, axially
symmetric material, similar to the phenomena seen during an outburst in
Wirtanen on 2018~September~26 \citep{FarnhamEtal:Wirt_tess}, that is removed
in our enhancements.

We explored additional aspects of these outbursts using our Monte-Carlo
model, with results presented by \cite{KnightEtal:wirtanen}.
\citet{KelleyEtal:wirt_outb} also provide additional analyses.

\section{Discussion and Summary}

\subsection{CN Coma Morphology \label{sec:disc_morph}}

The basic morphology revealed in our enhanced CN observations consistently
shows two jets that were tracked throughout the apparition.  One of the jets
remained active over most of a rotation, while the other appeared to turn on
and off with rotation. Although we can't conclude that the two jets arose
from the same active areas throughout the apparition, the sources always
appeared roughly half a rotation apart suggesting that they could be the
same.  Early and late in our observations, the jets produced spirals around
the nucleus, but they rotated clockwise pre-perihelion and counterclockwise
post-perihelion, suggesting that the Earth crossed the comet's equator around
perihelion.  \cite{SamarEtal:wirtanen} originally argued that the nucleus of
comet Wirtanen was likely in a non-principal axis (NPA) rotation state, but
later work \citep[e.g.,][]{GutierEtal:outgas} suggested that the angular
momentum might change, without resulting in an excited rotation state.
Indeed, we see no evidence for NPA rotation in our data.  Previous morphology
analyses \citep[e.g.,][]{Knight_Schl:hartley2, SamarEtal:Hartley2} have shown
that coma morphology can be a powerful tool for revealing evidence of NPA
states, but the features in our image sequences, when phased to the relevant
period, remain consistent from one rotation to the next and from one night to
the next. Thus, we conclude that Wirtanen's nucleus is in a near-simple
rotation state.  Our measured CN velocities, 0.62~km~s$^{-1}$ in mid-November
($r_H$=1.14~au) and 0.80~km~s$^{-1}$ in mid-January (1.13~au) are consistent
with other measurements of gas outflow for comets with similar gas production
and heliocentric distance \citep[e.g.,][]{TsengEtal:gas_vel, LeeEtal:miro}.

We used our CN morphology to constrain Monte-Carlo models of the coma,
to derive additional characteristics of the comet.  This work is presented by
\cite{KnightEtal:wirtanen}, and the results that the two studies have in
common are in general agreement.

\subsection{CN Lightcurves \label{sec:disc_lc}}

Our lightcurves are all double peaked, suggesting that there are two primary
active areas on the nucleus.  (Unlike in asteroid lightcurves, where a double
peaked lightcurve represents the changing cross-sectional surface area of a
spinning body, the variations in our coma lightcurves are produced by the
changing production rates of active areas as they rotate into and out of
sunlight.)  The phase separation ($\sim$0.52 from the primary to
the secondary peak) suggests the two active areas are separated by an
effective longitude $\sim$190\degr\ (or $\sim$170\degr\ in the other
direction).  This two-source configuration is consistent with the structures
we see in the morphology.

Comparing our phased lightcurves to the enhanced morphology images helps in
the interpretation of the lightcurve details.  The timing of a peak typically
occurs $\sim$0.1 phase after the initial appearance of a jet in the
morphology, with the brighter peak matching the jet that remains active
throughout the full rotation.  The phase offsets are the result of the delay
between the start of emission and the point at which material exits the
aperture.  This timing changes somewhat over the course of the apparition,
because of the changing dimensions of the 10-arcsec aperture at the comet.
The changing shape of the lightcurve and relative brightnesses of the peaks
is caused by the viewpoint evolution (where the spirals early and late exit
the aperture more rapidly than the face-on material around close approach)
combined with changes in the relative production rates of the two sources.
These same effects are likely the cause of the variations in the phase
separation between the high and low peak (0.50 to 0.54).

Although the relevant dates are not absolutely calibrated, we do detect
evidence of the two outbursts in our lightcurve measurements.  There are only
three sets of measurements, spanning half a rotation, from December~12.  They
seem to match well with the phased lightcurve, but the entire night requires
an offset of $\sim$0.1~mag --- significantly higher than any other night of
comparable quality --- to bring the data into line with the following nights.
This shows that the CN dominates the variability of the lightcurve, but has
underlying continuum from the outburst that systematically increases the
brightness.  The data quality on January~28 is lower, with more scatter at
various times during the night, but the outburst still affects the
lightcurve.  As noted in Section~\ref{sec:outbursts}, the first few hours of
photometry are $\sim$0.03~mag brighter than the same phase captured later in
the night, suggesting that we see some contribution from the outburst.

\subsection{Rotation Period \label{sec:disc_period}}

We used three different techniques to measure Wirtanen's apparent rotation
period at different epochs.  All of these techniques are in excellent
agreement and show that the period increased from our initial measurement of
8.98~hr in November to 9.14~hr at perihelion. After perihelion, it decreased
again, reaching our final measurement of 8.94~hr in February.  Measurements
from TRAPPIST telescopes spanning 12.5~hr on 2018~December~9-10 showed a
period 9.19$\pm$0.05~hr \citep{JehinEtal:cbet4585, MoulaneEtal:wirtanen},
which agrees with our results to within the errorbars.  Two other
measurements of Wirtanen's rotation period exist, both measured from sparse
data sets obtained in 1996.  \cite{MeechEtal:wirt} report a 7.6~hr period
from 1996~August~17/18 (-209 day from perihelion), and
\cite{LamyEtal:wirthst} report a period of 6.0$\pm$0.3~hr from 1996~August~28
(-198~day).  These results are discussed further below.

Our measurements represent apparent periods, but we showed that synodic
effects are too small to produce the observed changes, and thus conclude that
the nucleus is indeed changing its rotation rate with time.  This indicates
that the comet's activity is producing significant torques, with the net
direction of those torques changing direction around the time of perihelion.
Our observations suggest there are two primary active areas, separated by
$\sim$170\degr.  It is not difficult to conceive of scenarios, given a
non-spherical nucleus, where seasonal effects change the relative levels of
activity of the sources, reversing the direction of the torque.  See our
companion paper \citep{KnightEtal:wirtanen} for additional exploration of
this issue.  Similar behavior was seen in Rosetta measurements, where comet
67P/Churyumov-Gerasimenko (C-G) was observed to initially increase its rotation
period by $\sim$0.03~hr before the net torques reversed direction and
decreased the period by $\sim$0.37~hr \citep{KellerEtal:67P_period,
  KramerEtal:67P_rot}.

The trends in our measurements suggest that the period is already increasing
in November, and continues to decrease in February, thus neither of these
measurements represents an end state to the period changes, though the
February measurement, at +57~day, already hints that the rate of change is
slowing.  If we assume that the torques act in proportion to water
production, then by the end of March ($\sim$+100~day), where production rates
are $<$10\% of their peak value \citep{KnightEtal:wirtanen}, the torques
should largely be gone.  As both the production rates and the period changes
in Wirtanen appear to be symmetric around perihelion (ignoring synodic
effects), we can further assume that the period began increasing at -100~day
(early September).  It is notable that the window within 100~day of
perihelion is also where C-G exhibited the bulk of its changes
\citep{KramerEtal:67P_rot}.  If we extrapolate the ends of the curve in
Figure~\ref{fig:periods} so they flatten out at $\pm$100~days, then the period
at these extremes is likely to be around 8.8~hr.  Thus, the period exhibits a
maximum excursion of $\sim$4\% around perihelion, but it has nearly the same
value at both the start and end of the apparition.

With this in mind, we can evaluate the periods measured in 1996.  The 7.6~and
6.0~hr periods \citep{MeechEtal:wirt,LamyEtal:wirthst} represent changes of
16\% and 45\%, respectively, from our starting period, over four apparitions.
Barring any major alterations in the comet's torques due to activity or pole
orientation, neither of these measurements is consistent with our result.
Given that they also disagree with each other, though they were obtained only
11 days apart, we suggest that at least one of the results (most likely
the 7.6-hr ground-based measurement) may have contained more coma
contamination than assumed, and thus produced faulty results.  As noted,
the general trends that we see in 2018 are insufficient to alter the period
from 6~hr to 9~hr in four apparitions, but we cannot preclude unusual
activity in the intervening years that could have produced these changes,
and there is evidence for potentially significant outburst activity during
the 2002 apparition \citep{CombiEtal:water_prod} that perhaps could account for
changes on this scale.

The 2018/2019 apparition of comet Wirtanen, with an historically close
approach to the Earth, provided an excellent opportunity for studying this
important comet in detail.  We used this opportunity to acquire a wealth of
observations using both broadband and narrowband data, from which we derived
a number of important characteristics of the comet.  Our first results are
presented in this document and in our companion paper by
\cite{KnightEtal:wirtanen} and we expect to continue our analyses in the
future.


\acknowledgments

We would like to thank Annika Gustafsson, Michael Kelley, Uwe Konopka, Nick
Moskovitz, and Larry Wasserman for helping to obtain observations. These
results made use of the Lowell Discovery Telescope (LDT) at Lowell
Observatory. Lowell is a private, non-profit institution dedicated to
astrophysical research and public appreciation of astronomy and operates the
LDT in partnership with Boston University, the University of Maryland, the
University of Toledo, Northern Arizona University and Yale University.  The
Large Monolithic Imager was built by Lowell Observatory using funds provided
by the National Science Foundation (AST-1005313). This work also made use of
the Lowell 42-inch Hall Telescope, and the Lowell 31-inch telescope in
partnership with the University of Maryland.  Support was provided by NASA
Grants 80NSSC18K1007, 80NSSC18K0856, HST-GO-15372.006, and NSF 1852589.

%

\vspace{5mm}
\facilities{Lowell Discovery Telescope, Lowell Observatory 42-inch (1.1-m)
  Hall Telescope,  Lowell Observatory 31-inch (0.8-m) Telescope}


\software{IDL
          }

\end{document}